\documentclass[prd,aps,twocolumn,superscriptaddress,nofootinbib,preprintnumbers,floatfix]{revtex4-1}
\pdfoutput=1

\usepackage{amsmath}
\usepackage{graphicx}
\usepackage[small]{subfigure}
\usepackage{hyperref}
\usepackage{xspace}
\usepackage{xcolor}

\newcommand{\eq}[1]{Eq.~\eqref{eq:#1}}

\newcommand{\fig}[1]{Fig.~\ref{fig:#1}}

\newcommand{\abs}[1]{\lvert#1\rvert}
\newcommand{\Abs}[1]{\bigl\lvert#1\bigr\rvert}
\newcommand{\ord}[1]{\mathcal{O}(#1)}

\newcommand{\df}{\mathrm{d}}
\newcommand{\img}{\mathrm{i}}

\newcommand{\cB}{\mathcal{B}}
\newcommand{\cL}{\mathcal{L}}

\newcommand{\GeV}{\,\mathrm{GeV}}
\newcommand{\TeV}{\,\mathrm{TeV}}

\newcommand{\nn}{\nonumber}

\newcommand{\bn}{{\bar{n}}}
\newcommand{\as}{\alpha_s}

\newcommand{\mX}{m_X}
\newcommand{\pTjet}{p_T^{\rm jet}}
\newcommand{\pTcut}{p_T^{\rm cut}}
\newcommand{\nons}{\mathrm{nons}}
\newcommand{\F}{\mathcal{F}}

\begin{document}


\preprint{\vbox{
\hbox{DESY 16-086}
\hbox{MIT--CTP 4804}
\hbox{NIKHEF 2016-019}
}}

\title{Exploiting jet binning to identify the initial state of high-mass resonances}

\author{Markus A.~Ebert}
\author{Stefan Liebler}
\affiliation{Deutsches Elektronen-Synchrotron (DESY), D-22607 Hamburg, Germany}

\author{Ian Moult}
\author{Iain W.~Stewart}
\affiliation{Center for Theoretical Physics, Massachusetts Institute of Technology, Cambridge, Massachusetts 02139, USA}

\author{Frank J.~Tackmann}
\author{Kerstin Tackmann}
\affiliation{Deutsches Elektronen-Synchrotron (DESY), D-22607 Hamburg, Germany}

\author{Lisa Zeune}
\affiliation{Nikhef, Theory Group, Science Park 105, 1098 XG, Amsterdam, The Netherlands}

\date{May 19, 2016}

\begin{abstract}

If a new high-mass resonance is discovered at the Large Hadron Collider, model-independent techniques to
identify the production mechanism will be crucial to understand its nature and effective couplings to Standard Model particles.
We present a powerful and model-independent method to infer the initial state in the production
of any high-mass color-singlet system by using a tight veto on accompanying hadronic jets
to divide the data into two mutually exclusive event samples (jet bins).
For a resonance of several hundred GeV, the jet binning cut needed to discriminate quark and gluon initial states
is in the experimentally accessible range of several tens of GeV.
It also yields comparable cross sections for both bins, making this method viable
already with the small event samples available shortly after a discovery.
Theoretically, the method is made feasible by utilizing an effective field theory setup
to compute the jet cut dependence precisely and model independently and
to systematically control all sources of theoretical uncertainties in the jet binning, as well as their correlations.
We use a $750$\,GeV scalar resonance as an example to demonstrate
the viability of our method.

\end{abstract}

\maketitle

\section{Introduction}

The increased center-of-mass energy of the Large Hadron Collider (LHC)
significantly enhances the sensitivity for the discovery of new heavy particles.
Should a new high-mass state be found,
a key goal will be to identify its production mechanism.

It is well known that the different patterns of initial-state radiation (ISR) for
gluon- and quark-induced processes provide in principle a way to discriminate
between these initial states. Typically, methods to exploit this fact require a
substantial amount of data for the precise measurement of shapes of differential
distributions. In this paper, we show that for any high-mass color-singlet system,
the measurement of just two cross sections, namely dividing the data into events
with and without additional hadronic jets in the final state, provides a strong
discrimination between production mechanisms, which is furthermore experimentally
accessible with event samples of limited size.
The method is also theoretically clean, as it is both model independent and has
well-controlled theory uncertainties.

As a concrete example, we investigate a color-singlet resonance with a mass of $750$\,GeV.
The ATLAS and CMS experiments have recently reported some deviation
from the background expectation in the diphoton invariant mass spectrum around $750$ GeV~\cite{ATLAS750, CMS750}.
Assuming the deviation to be a first sign of a new particle, a large number of proposals on its interpretation and possible property studies have been made~\cite{diphoton:all}. Exploratory studies of the initial state have utilized the luminosity ratio between $8$ and $13$ TeV (which is limited by the available 8 TeV data), the transverse momentum and rapidity distribution of the new state~\cite{Gao:2015igz}, multiplicity and kinematic distributions of hadronic jets~\cite{Bernon:2016dow}, and $b$-tagging~\cite{Gao:2015igz, Franceschini:2016gxv}.
Different techniques for tagging the initial state have been studied earlier, see e.g.~Refs.~\cite{Sung:2009iq,Papaefstathiou:2009hp,Papaefstathiou:2010ru,Cox:2010ug, Krohn:2011zp,Ask:2011zs}.
Beyond its viability for small data sets, our method offers several additional advantages.
Compared to considering additional jets at high $\pTjet$~\cite{Grojean:2013nya, Bernon:2016dow,Franceschini:2016gxv},
the low $\pTjet$-range we exploit has more discrimination power and is more model independent.
Compared to the diphoton $p_T$ spectrum, the $\pTjet$ of hadronic jets provides a more direct measure of ISR, making it insensitive to the possibility of more complicated decays of the resonance, for example, three-body~\cite{Bernon:2015abk, An:2015cgp} or cascade decays \cite{Knapen:2015dap, Kim:2015ron, Cho:2015nxy, Altmannshofer:2015xfo, Liu:2015yec, Franceschini:2015kwy}.
Our method is also unaffected by limited experimental acceptance for photons, which for example hinders fully exploiting
the diphoton rapidity distribution to discriminate valence quarks by their different parton distribution function (PDF) shapes.

The 0-jet cross section is defined by requiring that all accompanying jets have $\pTjet\le \pTcut$.
The QCD dynamics of low-$p_T$ radiation produced in association with a hard scattering process into a final state $F$ with total invariant mass $m(F) \simeq \mX$ can be described using the soft collinear effective theory (SCET)~\cite{Bauer:2000ew, Bauer:2000yr, Bauer:2001ct, Bauer:2001yt}. At the scale $\mu\sim \pTcut\ll m_X$, the leading effective field theory (EFT) Lagrangian has the form (see e.g.~\cite{Stewart:2009yx, Berger:2010xi, Moult:2015aoa})
\begin{align} \label{eq:LeffpTjet}
\cL_\mathrm{eff}(\pTcut) 
  &= \cL_{\rm SCET} 
  +  c_{ggF}^{\lambda_1\lambda_2}\, \cB_n^{\lambda_1} \cB_\bn^{\lambda_2}\,\F
  \\ \nn & \quad
  +  \sum_q c_{q\bar qF}^{\lambda_1\lambda_2}\, \bar\chi_{q n}^{\lambda_1} \chi_{q \bn}^{\lambda_2}\,\F
\,.\end{align}
Here, $\cL_{\rm SCET}$ is the universal Lagrangian encoding the interactions of soft and collinear quarks and gluons.
The gauge-invariant operators $\cB_n \cB_\bn$ and $\bar\chi_{q n} \chi_{q \bn}$ describe the annihilation of energetic gluons or quarks $q = u,d,s,c,b$ along the beam directions, $n = (1, \hat z)$ and $\bn = (1,-\hat z)$, with helicities $\lambda_1$ and $\lambda_2$ (implicitly summed over), and $\F$ collects all fields required to produce $F$. All hard degrees of freedom are integrated out, including quarks and gluons of virtuality $\sim \mX$ as well as any intermediate new heavy degrees of freedom leading to $F$.

We stress that $\cL_\mathrm{eff}(\pTcut)$ provides a \emph{completely} model-independent description of the small-$\pTcut$ region, up to power corrections suppressed by $(\pTcut/\mX)^2$.
It is valid for \emph{any} produced color-singlet system $X$ leading to $F$, for example the decay of a finite-width resonance, the Standard Model background $pp\to F$, and even the signal-background interference~\cite{Moult:2014pja},
as all of the dynamics of the production and decay of $X$ are contained in the hard Wilson coefficients $c_{ggF}^{\lambda_1\lambda_2}$ and $c_{q\bar qF}^{\lambda_1\lambda_2}$ and because
the leading perturbative SCET dynamics are insensitive to the helicity structure of the operators and the details of $\F$. The 0-jet cross section thus only depends on the hard coefficients
\begin{equation} \label{eq:CiiX}
\Abs{c_{ijF}}^2 = \int\!\df\Phi_F \sum_{\lambda_1 \lambda_2}\! \Abs{c_{ijF}^{\lambda_1\lambda_2}(\Phi_F)}^2
\,,\end{equation}
where the integral is over the final-state phase space for $F$ including any kinematic selection cuts.
As a result, the $\pTcut$ dependence is independent of any details of $F$ and in particular also the spin of $X$.
(See Ref.~\cite{Tackmann:2016jyb} for a detailed analysis in a specific new-physics context.)

To predict the inclusive cross section $pp\to X\to F$ we need to know the Lagrangian at the scale $\mu \sim \mX$, which contains the full QCD Lagrangian plus the (effective) interactions of $X$ with quarks and gluons. This becomes somewhat more model dependent, and requires, for example, specifying the spin of $X$. For our concrete study we take $X$ to be a scalar, coupling to gluons and quarks via the effective Wilson coefficients $C_g$ and $C_q$ as
\begin{equation} \label{eq:LeffmX}
\cL_{\rm eff}(\mX) \supset -\frac{C_g}{1\,{\rm TeV}} \as G^{\mu\nu} G_{\mu\nu}X - \sum_q C_q\, \bar q q\, X
\,,\end{equation}
where $G^{\mu\nu}$ is the gluon field strength and $\as$ is the strong coupling.
(We assume that $X$ does not couple directly to top quarks, as this would have shown up in
$t\bar{t}$ production.)
Comparing quark and gluon luminosities as is often done is equivalent to using \eq{LeffmX} at leading order (LO).
With \eq{LeffmX} specified, we can now match it onto \eq{LeffpTjet} and compute $\abs{c_{q\bar qF}}^2$
and $\abs{c_{ggF}}^2$.

For our purposes, any model can be represented by \eq{LeffmX} at leading order in $\as(m_X)$.
Treated as an EFT, \eq{LeffmX} is \textit{a priori} only correct to $\ord{\mX/\Lambda}$, where $\Lambda$ is
the mass scale of additional heavy degrees of freedom that induce the effective interactions of $X$.
For example, the possibility of real QCD radiation from internal heavy states is not captured by 
\eq{LeffmX}. However, even for $\Lambda\sim\mX$ hard emissions only affect the inclusive cross section by $\ord{\as(m_X)}$,
while emissions below the scale $\pTcut$ are power suppressed.
Similarly, a different choice of $\cL_{\rm eff}(\mX)$ in \eq{LeffmX} (e.g.\ for a spin-2 resonance) changes the inclusive
cross section and the matching in \eq{matchmX} only by terms of $\ord{\as(m_X)}$, i.e., at the 10\%-20\% level.
The crucial point is that the $\pTcut$ dependence for $\pTcut\ll \mX$ is described by \eq{LeffpTjet}. Hence,
for more complicated scenarios than the one considered here,  our
main conclusions regarding the initial-state discrimination are unaffected as they 
rest on the dynamics at the scale $\mu \sim \pTcut$, which is described model independently.

\section{Calculational Setup}

Considering for simplicity the narrow-width approximation, we have
\begin{align} \label{eq:matchmX}
\Abs{c_{q\bar qF}(\mu_H)}^2 &= \cB(X\to F)\, \Abs{C_q(\mu_H)\, (1 + \dotsb)}^2
\,,\\\nn
\Abs{c_{ggF}(\mu_H)}^2 &= \cB(X\to F)\, \Abs{ \as(\mu_H)\, C_g(\mu_H)\, (1 +  \dotsb) }^2
\,,\end{align}
where $\mu_H\sim \mX$ is the hard matching scale, and the ellipses indicate the $\as(\mu_H)$ corrections
from hard virtual QCD emissions.
The branching ratio $\mathcal{B}\equiv \mathcal{B}(X\to F)$ also
depends on all Wilson coefficients $C_i$, but will drop out in our final analysis.

The jet cross sections we consider are given by
\begin{align} \label{eq:sigmaij}
\sigma_{\geq 0} &= |C_g|^2 \sigma_{\geq 0}^g + \sum_q |C_q|^2\sigma_{\geq 0}^q
\,,\nn\\
\sigma_0(\pTcut) &= |C_g|^2 \sigma_0^g(\pTcut) + \sum_q |C_q|^2\sigma_0^q(\pTcut)
\,,\nn\\
\sigma_{\geq 1}(\pTcut) &= |C_g|^2 \sigma_{\geq 1}^g(\pTcut) + \sum_q |C_q|^2\sigma_{\geq 1}^q(\pTcut)
\,,\end{align}
where $\sigma_{\geq 0} = \sigma_0(\pTcut) + \sigma_{\geq 1}(\pTcut)$.
We take $C_i \equiv C_i(\mX)$ as the unknown parameters to be determined from the data. Their evolution from the fixed input scale $\mX$ to the hard matching scale $\mu_H$ is included in the $\sigma_m^i$ in \eq{sigmaij}, so they are defined to be scale independent to all orders.
The $\sigma_{\geq 0}^g$ and $\sigma_{\geq 0}^q$ are the inclusive cross sections that
follow from the $ggX$ and $q\bar q X$ operators in $\cL_{\rm eff}(\mX)$ in \eq{LeffmX}. The inclusive $1$-jet cross sections are computed as $\sigma^i_{\geq 1}(\pTcut) = \sigma^i_{\geq 0} - \sigma^i_0(\pTcut)$.
The $0$-jet cross sections contain large Sudakov logarithms of $\pTcut/\mX$, which are resummed utilizing the $\pTjet$ resummation framework of Refs.~\cite{Tackmann:2012bt, Stewart:2013faa} based on SCET (see also Refs.~\cite{Banfi:2012jm, Becher:2012qa, Becher:2013xia, Banfi:2015pju}). They are given by
\begin{align} \label{eq:sigma0}
|C_g|^2 \sigma_0^g(\pTcut)
&= \frac{\pi}{4 E_\mathrm{cm}^2} \frac{m_X^2}{\TeV^2} \Abs{c_{ggX}(\mu)}^2
 \int\!{\mathrm{d}}Y   B_g(\pTcut, \mu) 
 \nn \\ &\quad 
 \times B_g(\pTcut, \mu)
 S_{gg}(\pTcut, \mu)
+ \sigma_0^{g\,\nons}(\pTcut)
\,,\nn \\
|C_q|^2 \sigma_0^q(\pTcut)
&= \frac{\pi}{6 E_\mathrm{cm}^2} \Abs{c_{q\bar qX}(\mu)}^2
 \int\!{\mathrm{d}}Y   B_q(\pTcut, \mu) 
 \\ &\quad 
 \times B_{\bar q}(\pTcut, \mu) S_{q\bar q}(\pTcut, \mu)
+ \sigma_0^{q\,\nons}(\pTcut) \nn
\,.\end{align}
The $B_i$ are quark and gluon beam functions, which describe the dynamics of collinear radiation along the beam directions.
In general, the type of the incoming parton is changed by both collinear PDF evolution
and fixed-order corrections, so beyond LO an operator in the Lagrangian receives contributions from all PDFs.
However, in the $0$-jet cross section, both of these effects only occur up to the scale $\pTcut$ and are contained in the beam functions. Above the scale $\pTcut$, the parton type of the initial state is uniquely defined and matches between the beam function and the operator in the Lagrangian~\cite{Stewart:2009yx}. Similarly, the dynamics of wide-angle soft radiation, described by the soft functions $S_{gg/q\bar q}$, is unique to the parton type and does not change it.
The fact that the jet veto freezes the initial-state parton type at the scale $\pTcut$ is what lends our method its strong discrimination power, as it provides a large energy range between $\pTcut$ and $\mX$ where the initial state evolves without changing its type.

We calculate $\sigma_{\geq 0}^q$ to next-to-leading order (NLO) in $\alpha_s$, and $\sigma_0^q(\pTcut)$ is resummed to NLL$'+$NLO order.
Due to the substantially larger uncertainties for gluons, we include the full next-to-next-to-leading order (NNLO) corrections for $\sigma_{\geq 0}^g$, and $\sigma_0^g(\pTcut)$ is resummed to NNLL$'+$NNLO~\cite{Stewart:2013faa}. The inclusive cross sections
are obtained with \textsc{SusHi 1.6.0}~\cite{Harlander:2012pb, Harlander:2016hcx, Harlander:2015xur, Harlander:2002wh, Harlander:2010cz}.

The nonsingular corrections $\sigma_0^{i\,\nons}(\pTcut)$ in \eq{sigma0} contain the power corrections starting at $(\pTcut/m_X)^2$. They ensure that $\sigma_0^i(\pTcut)$ smoothly matches onto $\sigma^i_{\ge 0}$ for large $\pTcut$, and are
correspondingly included to NLO for quarks and NNLO for gluons.
They are extracted from the fixed-order $\pTcut$ spectra predicted by \eq{LeffmX}, obtained from \textsc{SusHi} for $q\bar q X$ and \textsc{MCFM}~\cite{Campbell:1999ah, Campbell:2011bn} for $ggX$.

We perform a careful analysis of the different sources of theoretical uncertainties and their correlations for
$\sigma_{\geq 0}$, $\sigma_{0}$, and $\sigma_{\geq 1}$. The total theory covariance matrix
for all parton types and bins under consideration
is then obtained by adding the covariance matrices of all sources discussed below,
\begin{equation}
\label{eq:covth}
\mathcal{C}_{\rm th} = \mathcal{C}_{\rm FO}+\mathcal{C}_{\rm resum}+\mathcal{C}_{\varphi}+\mathcal{C}_{\rm PDF}
\,.\end{equation}
For the perturbative uncertainties we follow the treatment developed in
Refs.~\cite{Berger:2010xi, Stewart:2011cf, Stewart:2013faa} and
distinguish various independent sources. The first is an overall fixed-order
yield uncertainty, $\mathcal{C}_{\rm FO}$, which is fully correlated between all bins, 
and reproduces the usual fixed-order uncertainty for the inclusive cross section.
The resummation uncertainty, $\mathcal{C}_{\rm resum}$, is induced by the binning cut and
is correspondingly treated as a migration uncertainty that is fully anticorrelated between
$\sigma_0$ and $\sigma_{\geq 1}$ and drops out of $\sigma_{\geq 0}$.
The individual uncertainty contributions are estimated using profile
scale variations~\cite{Ligeti:2008ac,Abbate:2010xh} for the relevant
resummation scales,
as discussed in detail in Ref.~\cite{Stewart:2013faa}. Finally, we use a
complex hard scale $\mu_H = -\img \mX$ to resum large virtual QCD corrections.
The corresponding resummation uncertainty, $\mathcal{C}_{\varphi}$,
is estimated by varying the phase of $\mu_H$ and corresponds to a yield uncertainty.
The perturbative uncertainties are treated as fully correlated among all quark flavors and
uncorrelated between quarks and gluons.
We use the \textsc{MMHT2014nnlo68cl}~\cite{Harland-Lang:2014zoa} PDFs
with the corresponding $\as(m_Z) = 0.118$ and 3-loop running.
The parametric PDF uncertainties, $\mathcal{C}_{\rm PDF}$, are constructed from the $25$ independent eigenvectors
of \textsc{MMHT2014nnlo68cl}. They are subdominant compared to the perturbative uncertainties.

\section{Initial-state discrimination}

\begin{figure}[t]
\centering
\includegraphics[width=\columnwidth]{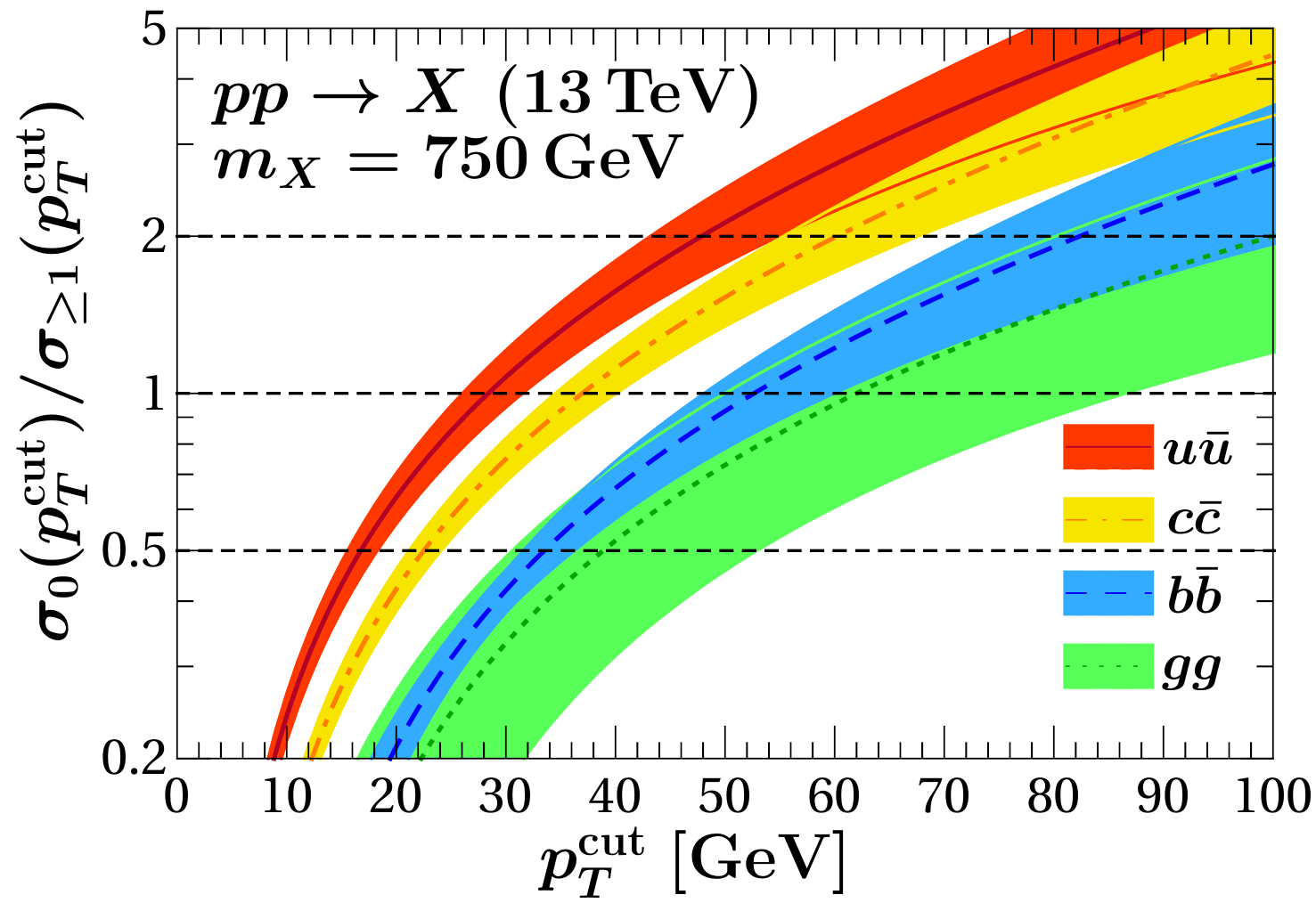}
\caption{The ratio $\sigma_{0}(\pTcut)/\sigma_{\geq 1}(\pTcut)$
for $u$ (red), $c$ (yellow), $b$ quarks (blue) and gluons (green).
The lines show the central values and the bands the theoretical uncertainties.
}
\label{fig:ratio0To1}
\end{figure}

\begin{figure*}[t]
\subfigure[]{\includegraphics[width=0.32\textwidth]{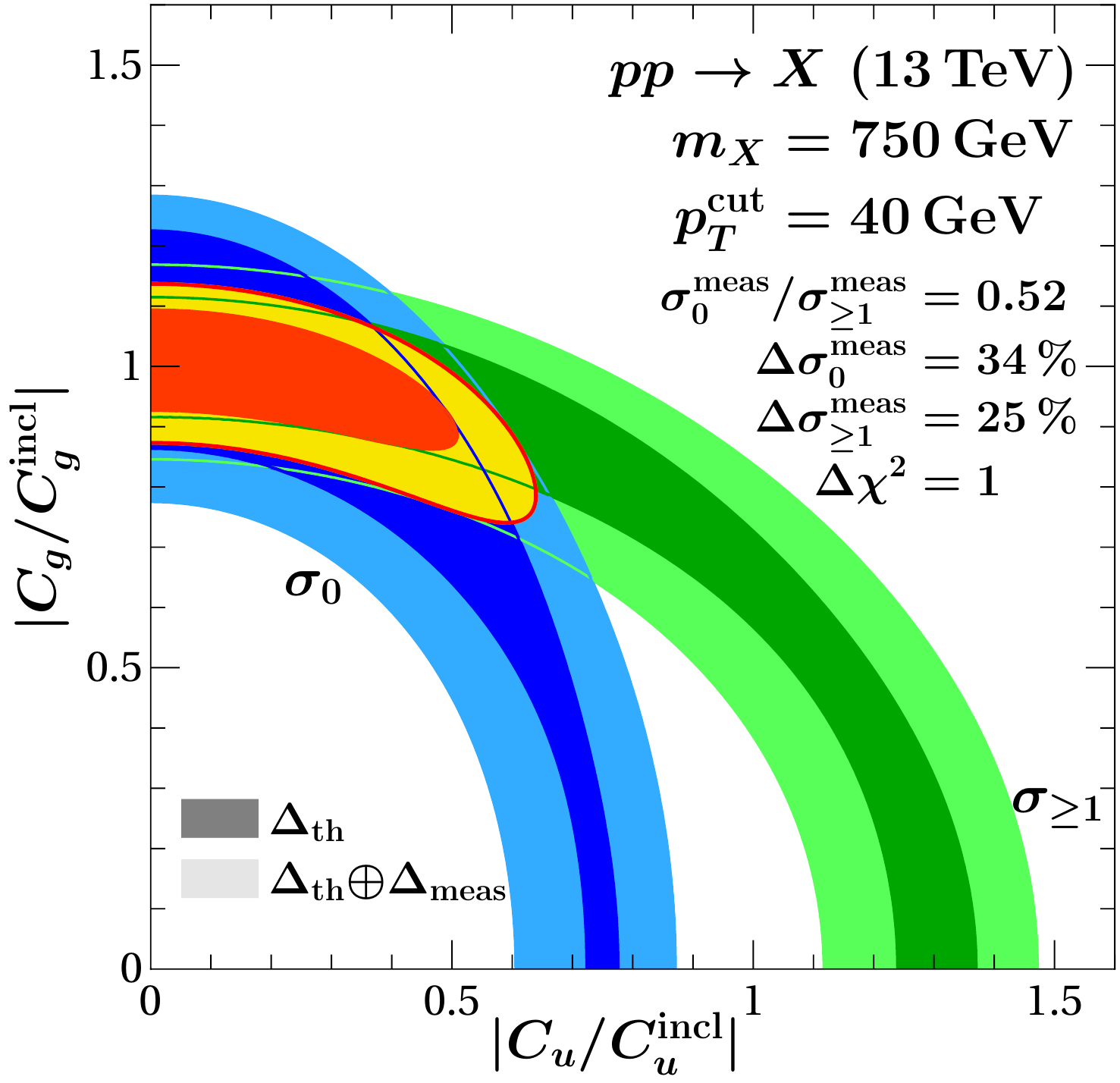}}%
\hfill%
\subfigure[]{\includegraphics[width=0.32\textwidth]{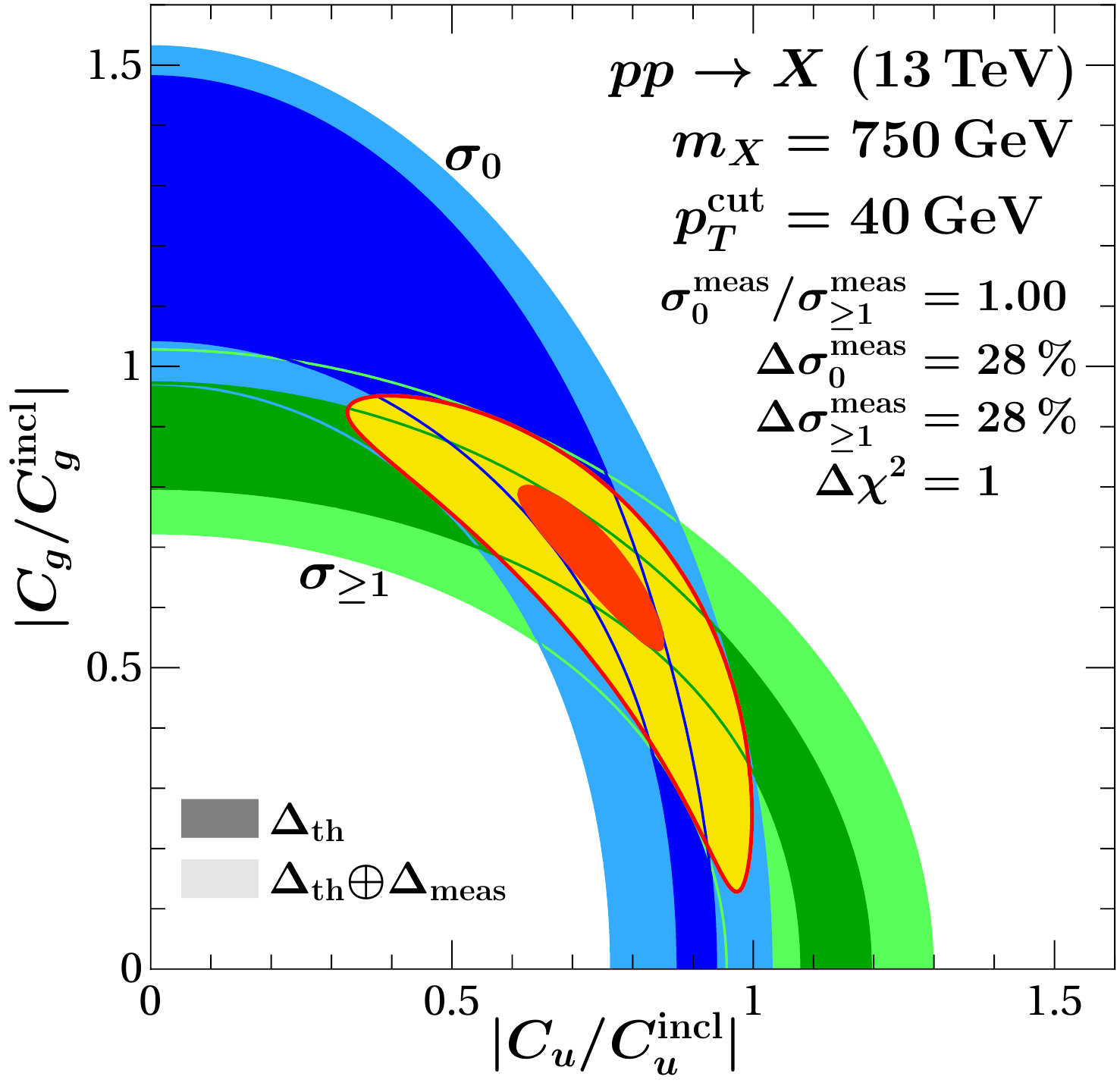}}%
\hfill%
\subfigure[]{\includegraphics[width=0.32\textwidth]{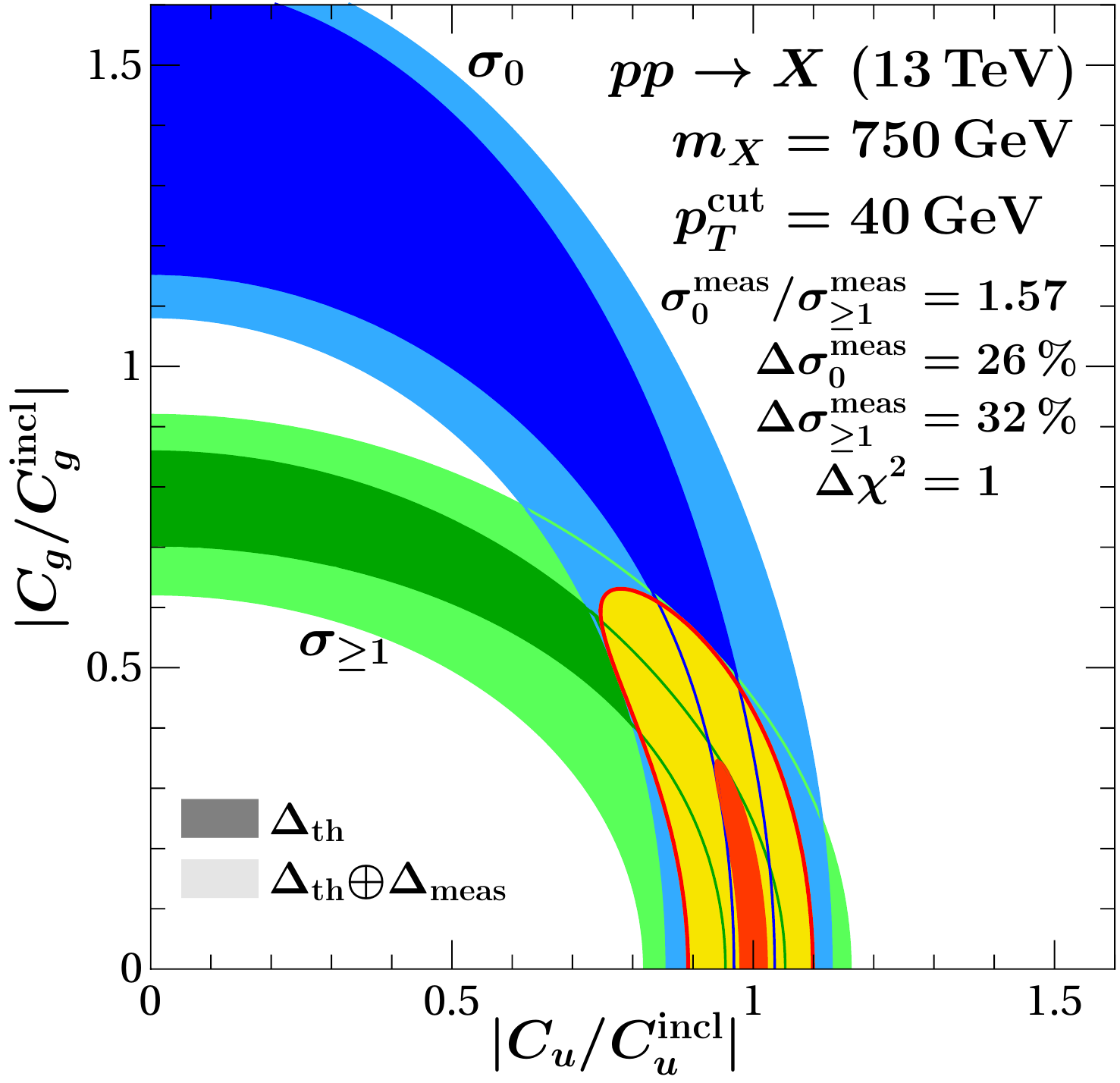}}%
\\[-2ex]
\subfigure[]{\includegraphics[width=0.32\textwidth]{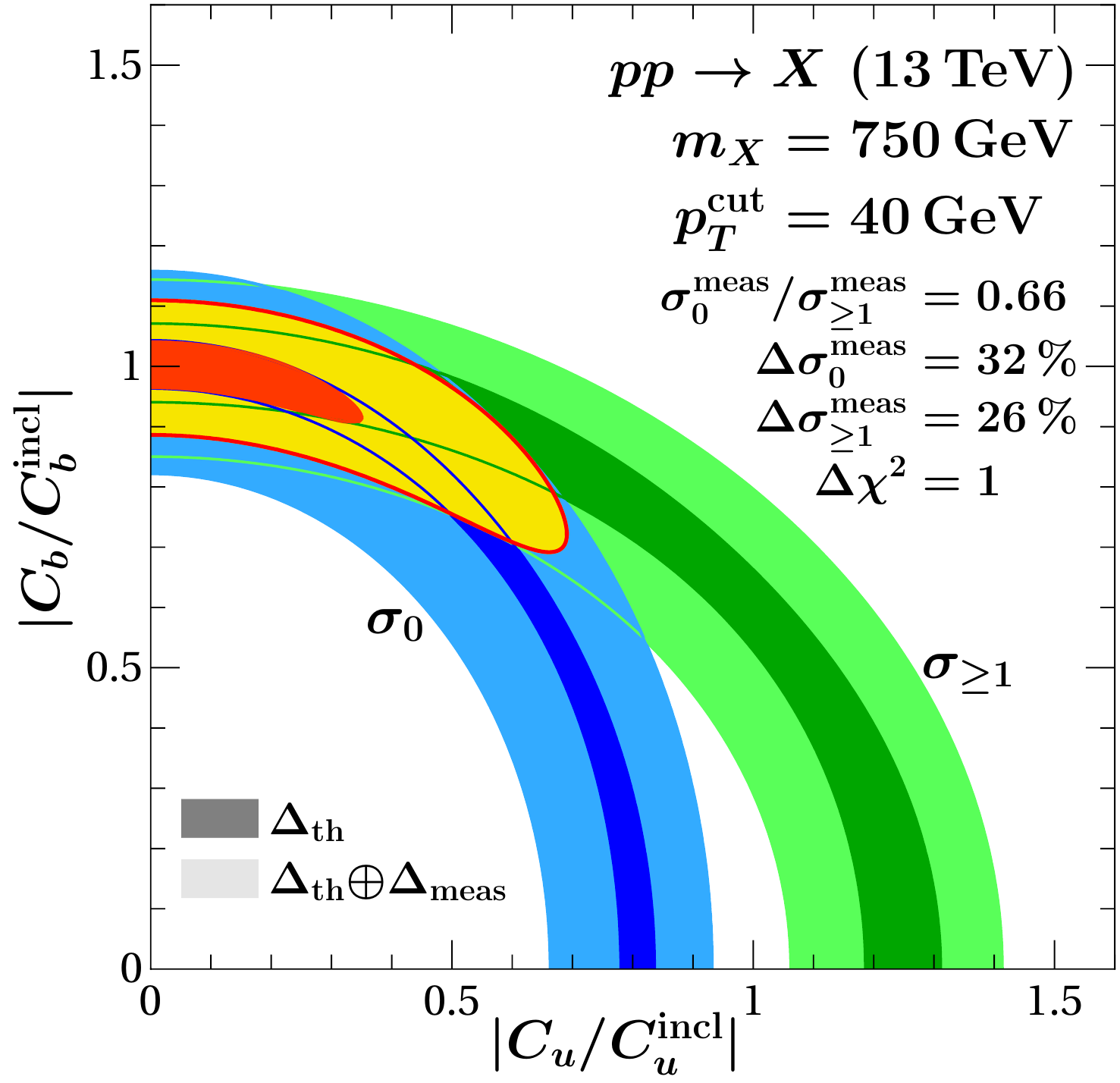}}%
\hfill%
\subfigure[]{\includegraphics[width=0.32\textwidth]{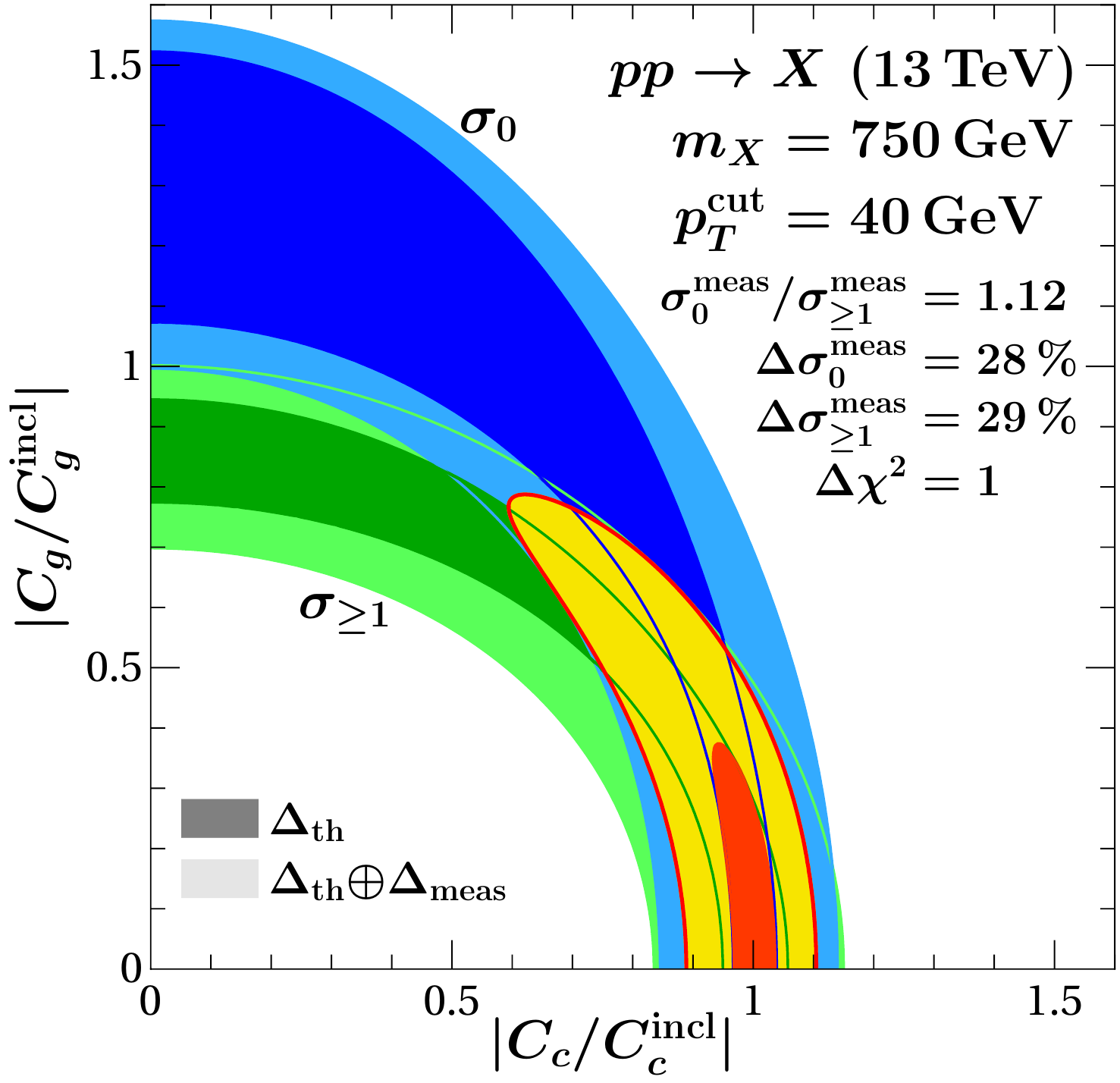}}%
\hfill%
\subfigure[]{\includegraphics[width=0.32\textwidth]{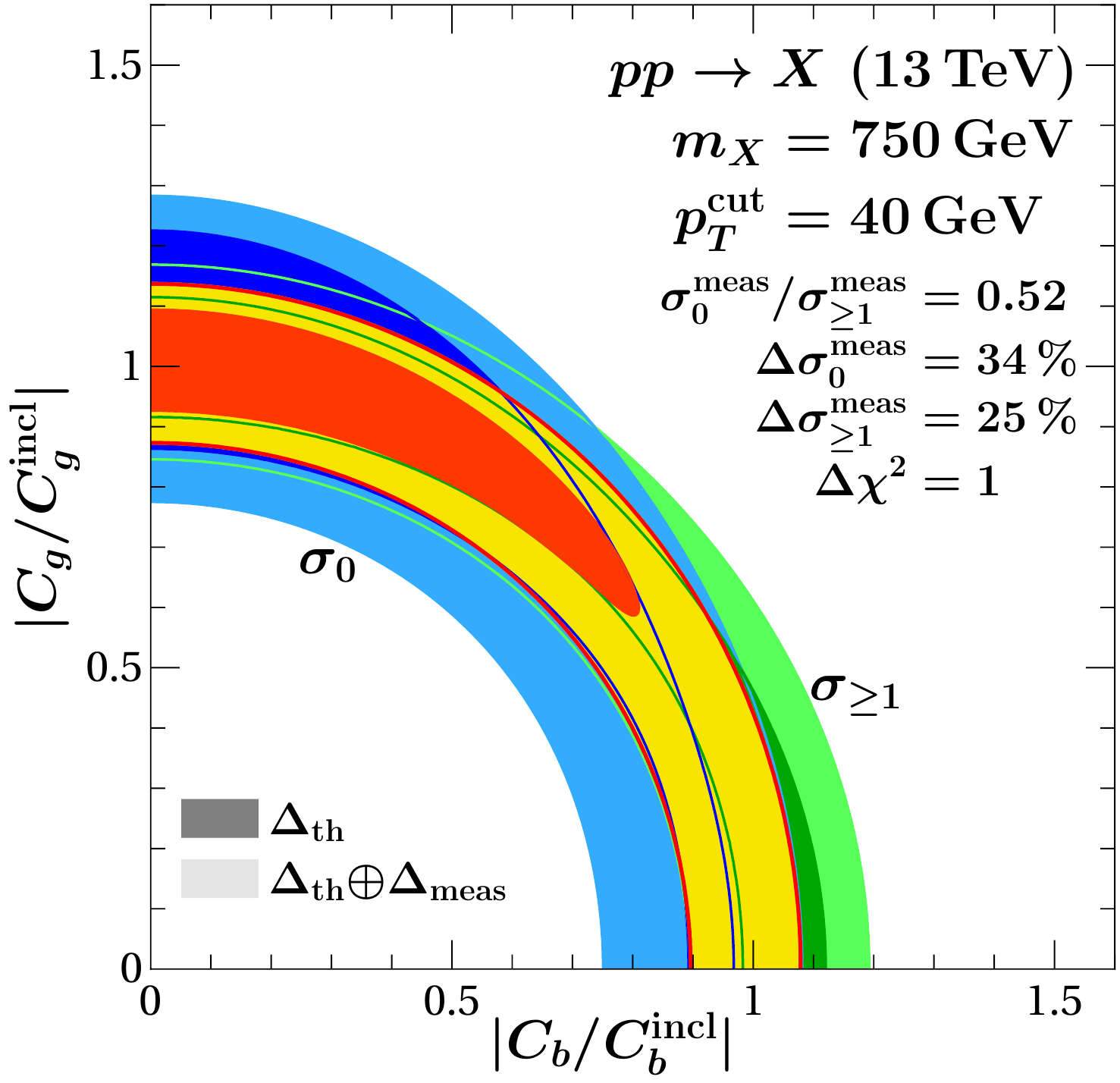}}%
\vspace{-2ex}
\caption{$\Delta\chi^2=1$-contours for various scenarios.
(a) gluon signal, (b) mixed gluon/$u$-quark signal, (c) $u$-quark signal, (d) $b$-quark signal, (e) $c$-quark signal, (f) gluon signal.
The constraints from $\sigma_{0}$ and $\sigma_{\ge1}$ are shown by the blue and green bands, respectively.
The combined constraint from both are shown by the orange/yellow regions.
The inner darker regions correspond to theory uncertainties only, while the full lighter bands include both theory and assumed experimental uncertainties.}
\label{fig:chi2}
\end{figure*}

As a demonstration of the technique, we consider a hypothetical scalar resonance of mass $\mX = 750\GeV$ produced in $13\TeV$ $pp$ collisions.
In \fig{ratio0To1} we show the ratio $\sigma^i_{0}(\pTcut)/\sigma^i_{\geq 1}(\pTcut)$
as a function of $\pTcut$ for different initial states $i = u,c,b,g$
including all theoretical uncertainties, see \eq{covth}.
For all our results we use a jet radius of $0.4$.
The analogous results for $d$ and $s$ quarks are mostly indistinguishable from $u$ quarks and are not shown.
The split of the cross section into the $0$-jet and ${\geq}1$-jet bins is clearly different for the different initial states,
allowing one to distinguish light quarks ($u$,$d$,$s$), $c$ quarks,
$b$ quarks, and gluons. The discrimination between $b$ quarks and gluons is less good,
in part due to the sizeable uncertainty in the gluon cross sections.

The optimal $\pTcut$ value for discriminating between different possible initial states depends on the true initial state.
Numerically, we observe only a mild sensitivity of the discrimination between initial states
on the $\pTcut$ value within the range $\pTcut\in[25,65]\GeV$.
The $\pTcut$ value can thus be chosen to optimize the experimental sensitivity with limited statistics.
A roughly equal split of the cross section is achieved at $\pTcut\simeq 25\GeV$ ($65\GeV$) for a light-quark (gluon) induced signal.
In our subsequent analysis we use $\pTcut = 40\GeV$, for which the cross section ratio is between 0.5 and 2 for any initial state.

Note that effects from hadronization and multiparton interactions, which are not included in our calculations,
can affect the leading jet $p_T$ spectrum at small $p_T$. However their effects partially compensate each other.
We checked that the net effect in the cross section ratios we consider becomes negligible
above $\pTcut \gtrsim 20\GeV$.

To study the constraints on the Wilson coefficients $C_i$ from measuring $\sigma_0$ and $\sigma_{\geq 1}$, we minimize
the $\chi^2$ function
\begin{equation} \label{eq:chi2func}
\chi^2(C_i) =\hspace{-3mm} \sum_{m,n \in \{ 0, {\ge}1 \}} \hspace{-3mm} (\sigma_m^{\mathrm{meas}}\!-\sigma_m)
(\mathcal{C}^{-1})_{mn} (\sigma_n^{\mathrm{meas}}\!-\sigma_n)
\,,\end{equation}
where $\sigma_m^{\mathrm{meas}}$ is the measured $pp\to X\to F$ cross section in bin $m$,
$\mathcal{C}$ is the sum of the experimental and theory covariance matrices,
and $\sigma_m$ is the predicted cross section in bin $m$ in \eq{sigmaij}.
In the narrow-width approximation and considering a single decay channel (e.g.~$F=\gamma\gamma$),
we only constrain $C_i \sqrt{\mathcal{B}}$ [see \eq{matchmX}].
To render our results independent of the details of $F$, we define
\begin{equation} \label{eq:cimax}
C_i^\mathrm{incl}\sqrt{\mathcal{B}} = \sqrt{ \sigma_{{\ge}0}^\mathrm{meas} / \sigma^i_{\ge0} }
\,,\end{equation}
which is the value of $C_i\sqrt{\mathcal{B}}$ for which the measured inclusive
cross section is completely attributed to initial state $i$.
By considering the ratios $|C_i/C_i^\mathrm{incl}|$, our analysis only depends
on the ratio $\sigma_0/\sigma_{\geq 1}$ but not the absolute cross sections or $\mathcal{B}$.
With more than one final state, the sum in \eq{chi2func} runs over the respective bins for all
final states, and the dependence of the different branching ratios
on the Wilson coefficients $C_i$ and the associated uncertainties need to be taken
into account.

Figure~\ref{fig:chi2} shows the constraints on $|C_i/C_i^\mathrm{incl}|$ that can be achieved for
various scenarios of the assumed true values of the $C_i$.
For this purpose, we assume that the inclusive cross section is measured
with a relative uncertainty of $20\%$, which can be realistically expected not very
long after a discovery of a new state. We split the cross section
into measured $0$-jet and ${\geq}1$-jet bins according to the
theoretical predictions for the assumed signal, with the resulting
$\sigma_{0}^\mathrm{meas}/\sigma_{\ge1}^\mathrm{meas}$ given in each plot.
The relative uncertainties, $\Delta \sigma_{m}^\mathrm{meas}$, on the measurements in the two bins
are assumed to be uncorrelated and split according to
$\Delta \sigma_{0}^\mathrm{meas}/\Delta \sigma_{\ge1}^\mathrm{meas}
=\sqrt{\sigma_{\ge1}^\mathrm{meas}/\sigma_{0}^\mathrm{meas}}$.

In \fig{chi2}, contours of $\Delta\chi^2= 1$ only including the theoretical
uncertainties are shown by the inner darker bands and combining theoretical and
assumed experimental uncertainties by the outer lighter bands. The individual constraints from $\sigma_0$
and $\sigma_{\geq 1}$ are shown by the blue and green bands, respectively, while the
combined constraint from both is shown in yellow/orange.

Figures~\ref{fig:chi2} (a)-(c) illustrates the good discrimination between light quarks, here $u$, and
gluons in the initial state, for a purely gluon-induced signal in (a), for a mixed signal with the cross section ratio equal to one in (b), and a purely $u$-quark induced signal in (c). Figures~\ref{fig:chi2} (d) and (e) demonstrate the good discrimination between $u$ and $b$ quarks for a $b$-quark signal, and between gluons and $c$ quarks for a $c$-quark signal, respectively.
Only the discrimination between $b$ quarks and gluons, shown in \fig{chi2} (f) for a gluon signal, remains challenging
due to the weaker separation already seen in \fig{ratio0To1}.

We conclude that even with fairly large experimental uncertainties, as expected soon after a potential discovery,
a clear separation between different initial states can be achieved for most scenarios.
A combined fit to all coefficients will of course require more data, and will then also benefit from using several $\pTcut$ values. We stress that thanks to the used resummation framework, the theoretical uncertainties and correlations can be robustly estimated and are not a limiting factor, and if necessary, could also be reduced further.

\section{Conclusions}

Should the deviation in the diphoton spectrum at $750\GeV$ manifest into a discovery, the method proposed here can be readily applied to identify its initial state. It is then preferable for
the measurements to be fiducial in the kinematics of the $X$ decay products to minimize the
model dependence introduced by acceptance corrections.

We restricted our attention to discriminate quark- and gluon-initiated production. Using our method, it will also be possible to identify photoproduction, which has been considered in several recent studies~\cite{Csaki:2015vek, Fichet:2015vvy, Harland-Lang:2016qjy, Abel:2016pyc, Martin:2016byf, Cynolter:2016jxv, Csaki:2016raa, Molinaro:2016oix}.
In this case, the $0$-jet cross section is given in terms of photon beam functions calculated in terms of photon PDFs at the scale $\mu \sim \pTcut$ and without QCD evolution above $\pTcut$. This implies that the $\sigma_0/\sigma_{\geq 1}$ ratio will be substantially larger than for light quarks, thus providing a good discrimination against the production via quarks and gluons.
The utility of a jet-veto to distinguish photon production from vector-boson fusion or gluon initial states was discussed in \cite{Harland-Lang:2016qjy}.

Our method allows for an early, model-independent, and theoretically clean identification of
the production mechanism of any new high-mass color-singlet state.
Since the ratio $\sigma^i_{0}(\pTcut)/\sigma^i_{\geq 1}(\pTcut)$ depends to good approximation only on $\pTcut / m_X$, and $\pTcut \gtrsim 25\GeV$ is experimentally feasible at the LHC, we expect it to work well for masses $m_X \gtrsim 300\GeV$.

\begin{acknowledgments}
This work was partially supported by the DFG Emmy-Noether Grant No. TA 867/1-1, the
Office of Nuclear Physics of the U.S. Department of Energy under the Grant No.~DE-SC0011090, the
Collaborative Research Center SFB 676 of the DFG ``Particles, Strings and the Early Universe'',
the Simons Foundation through the Investigator Grant 327942,
the Netherlands Organization for Scientific Research (NWO) through a VENI Grant,
the PIER Helmholtz Graduate School, and by a Global MISTI Collaboration Grant from MIT.
\end{acknowledgments}

\bibliography{diphoton750}

\begin{thebibliography}{55}%
\makeatletter
\providecommand \@ifxundefined [1]{%
 \@ifx{#1\undefined}
}%
\providecommand \@ifnum [1]{%
 \ifnum #1\expandafter \@firstoftwo
 \else \expandafter \@secondoftwo
 \fi
}%
\providecommand \@ifx [1]{%
 \ifx #1\expandafter \@firstoftwo
 \else \expandafter \@secondoftwo
 \fi
}%
\providecommand \natexlab [1]{#1}%
\providecommand \enquote  [1]{``#1''}%
\providecommand \bibnamefont  [1]{#1}%
\providecommand \bibfnamefont [1]{#1}%
\providecommand \citenamefont [1]{#1}%
\providecommand \href@noop [0]{\@secondoftwo}%
\providecommand \href [0]{\begingroup \@sanitize@url \@href}%
\providecommand \@href[1]{\@@startlink{#1}\@@href}%
\providecommand \@@href[1]{\endgroup#1\@@endlink}%
\providecommand \@sanitize@url [0]{\catcode `\\12\catcode `\$12\catcode
  `\&12\catcode `\#12\catcode `\^12\catcode `\_12\catcode `\%12\relax}%
\providecommand \@@startlink[1]{}%
\providecommand \@@endlink[0]{}%
\providecommand \url  [0]{\begingroup\@sanitize@url \@url }%
\providecommand \@url [1]{\endgroup\@href {#1}{\urlprefix }}%
\providecommand \urlprefix  [0]{URL }%
\providecommand \Eprint [0]{\href }%
\providecommand \doibase [0]{http://dx.doi.org/}%
\providecommand \selectlanguage [0]{\@gobble}%
\providecommand \bibinfo  [0]{\@secondoftwo}%
\providecommand \bibfield  [0]{\@secondoftwo}%
\providecommand \translation [1]{[#1]}%
\providecommand \BibitemOpen [0]{}%
\providecommand \bibitemStop [0]{}%
\providecommand \bibitemNoStop [0]{.\EOS\space}%
\providecommand \EOS [0]{\spacefactor3000\relax}%
\providecommand \BibitemShut  [1]{\csname bibitem#1\endcsname}%
\let\auto@bib@innerbib\@empty
\bibitem [{\citenamefont {{ATLAS Collaboration}}(2016)}]{ATLAS750}%
  \BibitemOpen
  \bibfield  {author} {\bibinfo {author} {\bibnamefont {{ATLAS
  Collaboration}}},\ }\href@noop {} {\  (\bibinfo {year} {2016})},\ \bibinfo
  {note} {{ATLAS-CONF-2016-018, ATLAS-CONF-2015-081}}\BibitemShut {NoStop}%
\bibitem [{\citenamefont {{CMS Collaboration}}(2016)}]{CMS750}%
  \BibitemOpen
  \bibfield  {author} {\bibinfo {author} {\bibnamefont {{CMS Collaboration}}},\
  }\href@noop {} {\  (\bibinfo {year} {2016})},\ \bibinfo {note}
  {{CMS-PAS-EXO-16-018, CMS-PAS-EXO-15-004}}\BibitemShut {NoStop}%
\bibitem [{dip()}]{diphoton:all}%
  \BibitemOpen
  \href@noop {} {}\bibinfo {howpublished} {A complete list of papers discussing
  Refs.~\cite{ATLAS750, CMS750} can be found at:
  \url{http://inspirehep.net/search?ln=en&p=refersto\%3Arecid\%3A1410174}}\BibitemShut
  {NoStop}%
\bibitem [{\citenamefont {Gao}\ \emph {et~al.}(2016)\citenamefont {Gao},
  \citenamefont {Zhang},\ and\ \citenamefont {Zhu}}]{Gao:2015igz}%
  \BibitemOpen
  \bibfield  {author} {\bibinfo {author} {\bibfnamefont {J.}~\bibnamefont
  {Gao}}, \bibinfo {author} {\bibfnamefont {H.}~\bibnamefont {Zhang}}, \ and\
  \bibinfo {author} {\bibfnamefont {H.~X.}\ \bibnamefont {Zhu}},\ }\href
  {\doibase 10.1140/epjc/s10052-016-4200-z} {\bibfield  {journal} {\bibinfo
  {journal} {Eur. Phys. J.}\ }\textbf {\bibinfo {volume} {C76}},\ \bibinfo
  {pages} {348} (\bibinfo {year} {2016})},\ \Eprint
  {http://arxiv.org/abs/1512.08478} {arXiv:1512.08478 [hep-ph]} \BibitemShut
  {NoStop}%
\bibitem [{\citenamefont {Bernon}\ \emph {et~al.}(2016)\citenamefont {Bernon},
  \citenamefont {Goudelis}, \citenamefont {Kraml}, \citenamefont {Mawatari},\
  and\ \citenamefont {Sengupta}}]{Bernon:2016dow}%
  \BibitemOpen
  \bibfield  {author} {\bibinfo {author} {\bibfnamefont {J.}~\bibnamefont
  {Bernon}}, \bibinfo {author} {\bibfnamefont {A.}~\bibnamefont {Goudelis}},
  \bibinfo {author} {\bibfnamefont {S.}~\bibnamefont {Kraml}}, \bibinfo
  {author} {\bibfnamefont {K.}~\bibnamefont {Mawatari}}, \ and\ \bibinfo
  {author} {\bibfnamefont {D.}~\bibnamefont {Sengupta}},\ }\href {\doibase
  10.1007/JHEP05(2016)128} {\bibfield  {journal} {\bibinfo  {journal} {JHEP}\
  }\textbf {\bibinfo {volume} {05}},\ \bibinfo {pages} {128} (\bibinfo {year}
  {2016})},\ \Eprint {http://arxiv.org/abs/1603.03421} {arXiv:1603.03421
  [hep-ph]} \BibitemShut {NoStop}%
\bibitem [{\citenamefont {Franceschini}\ \emph
  {et~al.}(2016{\natexlab{a}})\citenamefont {Franceschini}, \citenamefont
  {Giudice}, \citenamefont {Kamenik}, \citenamefont {McCullough}, \citenamefont
  {Riva}, \citenamefont {Strumia},\ and\ \citenamefont
  {Torre}}]{Franceschini:2016gxv}%
  \BibitemOpen
  \bibfield  {author} {\bibinfo {author} {\bibfnamefont {R.}~\bibnamefont
  {Franceschini}}, \bibinfo {author} {\bibfnamefont {G.~F.}\ \bibnamefont
  {Giudice}}, \bibinfo {author} {\bibfnamefont {J.~F.}\ \bibnamefont
  {Kamenik}}, \bibinfo {author} {\bibfnamefont {M.}~\bibnamefont {McCullough}},
  \bibinfo {author} {\bibfnamefont {F.}~\bibnamefont {Riva}}, \bibinfo {author}
  {\bibfnamefont {A.}~\bibnamefont {Strumia}}, \ and\ \bibinfo {author}
  {\bibfnamefont {R.}~\bibnamefont {Torre}},\ }\href {\doibase
  10.1007/JHEP07(2016)150} {\bibfield  {journal} {\bibinfo  {journal} {JHEP}\
  }\textbf {\bibinfo {volume} {07}},\ \bibinfo {pages} {150} (\bibinfo {year}
  {2016}{\natexlab{a}})},\ \Eprint {http://arxiv.org/abs/1604.06446}
  {arXiv:1604.06446 [hep-ph]} \BibitemShut {NoStop}%
\bibitem [{\citenamefont {Sung}(2009)}]{Sung:2009iq}%
  \BibitemOpen
  \bibfield  {author} {\bibinfo {author} {\bibfnamefont {I.}~\bibnamefont
  {Sung}},\ }\href {\doibase 10.1103/PhysRevD.80.094020} {\bibfield  {journal}
  {\bibinfo  {journal} {Phys. Rev. D}\ }\textbf {\bibinfo {volume} {80}},\
  \bibinfo {pages} {094020} (\bibinfo {year} {2009})},\ \Eprint
  {http://arxiv.org/abs/0908.3688} {arXiv:0908.3688} \BibitemShut {NoStop}%
\bibitem [{\citenamefont {Papaefstathiou}\ and\ \citenamefont
  {Webber}(2009)}]{Papaefstathiou:2009hp}%
  \BibitemOpen
  \bibfield  {author} {\bibinfo {author} {\bibfnamefont {A.}~\bibnamefont
  {Papaefstathiou}}\ and\ \bibinfo {author} {\bibfnamefont {B.}~\bibnamefont
  {Webber}},\ }\href {\doibase 10.1088/1126-6708/2009/06/069} {\bibfield
  {journal} {\bibinfo  {journal} {JHEP}\ }\textbf {\bibinfo {volume} {06}},\
  \bibinfo {pages} {069} (\bibinfo {year} {2009})},\ \Eprint
  {http://arxiv.org/abs/0903.2013} {arXiv:0903.2013} \BibitemShut {NoStop}%
\bibitem [{\citenamefont {Papaefstathiou}\ and\ \citenamefont
  {Webber}(2010)}]{Papaefstathiou:2010ru}%
  \BibitemOpen
  \bibfield  {author} {\bibinfo {author} {\bibfnamefont {A.}~\bibnamefont
  {Papaefstathiou}}\ and\ \bibinfo {author} {\bibfnamefont {B.}~\bibnamefont
  {Webber}},\ }\href {\doibase 10.1007/JHEP07(2010)018} {\bibfield  {journal}
  {\bibinfo  {journal} {JHEP}\ }\textbf {\bibinfo {volume} {07}},\ \bibinfo
  {pages} {018} (\bibinfo {year} {2010})},\ \Eprint
  {http://arxiv.org/abs/1004.4762} {arXiv:1004.4762} \BibitemShut {NoStop}%
\bibitem [{\citenamefont {Cox}\ \emph {et~al.}(2011)\citenamefont {Cox},
  \citenamefont {Forshaw},\ and\ \citenamefont {Pilkington}}]{Cox:2010ug}%
  \BibitemOpen
  \bibfield  {author} {\bibinfo {author} {\bibfnamefont {B.~E.}\ \bibnamefont
  {Cox}}, \bibinfo {author} {\bibfnamefont {J.~R.}\ \bibnamefont {Forshaw}}, \
  and\ \bibinfo {author} {\bibfnamefont {A.~D.}\ \bibnamefont {Pilkington}},\
  }\href {\doibase 10.1016/j.physletb.2010.12.011} {\bibfield  {journal}
  {\bibinfo  {journal} {Phys. Lett.}\ }\textbf {\bibinfo {volume} {B696}},\
  \bibinfo {pages} {87} (\bibinfo {year} {2011})},\ \Eprint
  {http://arxiv.org/abs/1006.0986} {arXiv:1006.0986} \BibitemShut {NoStop}%
\bibitem [{\citenamefont {Krohn}\ \emph {et~al.}(2011)\citenamefont {Krohn},
  \citenamefont {Randall},\ and\ \citenamefont {Wang}}]{Krohn:2011zp}%
  \BibitemOpen
  \bibfield  {author} {\bibinfo {author} {\bibfnamefont {D.}~\bibnamefont
  {Krohn}}, \bibinfo {author} {\bibfnamefont {L.}~\bibnamefont {Randall}}, \
  and\ \bibinfo {author} {\bibfnamefont {L.-T.}\ \bibnamefont {Wang}},\
  }\href@noop {} {\  (\bibinfo {year} {2011})},\ \Eprint
  {http://arxiv.org/abs/1101.0810} {arXiv:1101.0810} \BibitemShut {NoStop}%
\bibitem [{\citenamefont {Ask}\ \emph {et~al.}(2012)\citenamefont {Ask},
  \citenamefont {Collins}, \citenamefont {Forshaw}, \citenamefont {Joshi},\
  and\ \citenamefont {Pilkington}}]{Ask:2011zs}%
  \BibitemOpen
  \bibfield  {author} {\bibinfo {author} {\bibfnamefont {S.}~\bibnamefont
  {Ask}}, \bibinfo {author} {\bibfnamefont {J.~H.}\ \bibnamefont {Collins}},
  \bibinfo {author} {\bibfnamefont {J.~R.}\ \bibnamefont {Forshaw}}, \bibinfo
  {author} {\bibfnamefont {K.}~\bibnamefont {Joshi}}, \ and\ \bibinfo {author}
  {\bibfnamefont {A.~D.}\ \bibnamefont {Pilkington}},\ }\href {\doibase
  10.1007/JHEP01(2012)018} {\bibfield  {journal} {\bibinfo  {journal} {JHEP}\
  }\textbf {\bibinfo {volume} {01}},\ \bibinfo {pages} {018} (\bibinfo {year}
  {2012})},\ \Eprint {http://arxiv.org/abs/1108.2396} {arXiv:1108.2396}
  \BibitemShut {NoStop}%
\bibitem [{\citenamefont {Grojean}\ \emph {et~al.}(2014)\citenamefont
  {Grojean}, \citenamefont {Salvioni}, \citenamefont {Schlaffer},\ and\
  \citenamefont {Weiler}}]{Grojean:2013nya}%
  \BibitemOpen
  \bibfield  {author} {\bibinfo {author} {\bibfnamefont {C.}~\bibnamefont
  {Grojean}}, \bibinfo {author} {\bibfnamefont {E.}~\bibnamefont {Salvioni}},
  \bibinfo {author} {\bibfnamefont {M.}~\bibnamefont {Schlaffer}}, \ and\
  \bibinfo {author} {\bibfnamefont {A.}~\bibnamefont {Weiler}},\ }\href
  {\doibase 10.1007/JHEP05(2014)022} {\bibfield  {journal} {\bibinfo  {journal}
  {JHEP}\ }\textbf {\bibinfo {volume} {05}},\ \bibinfo {pages} {022} (\bibinfo
  {year} {2014})},\ \Eprint {http://arxiv.org/abs/1312.3317} {arXiv:1312.3317}
  \BibitemShut {NoStop}%
\bibitem [{\citenamefont {Bernon}\ and\ \citenamefont
  {Smith}(2016)}]{Bernon:2015abk}%
  \BibitemOpen
  \bibfield  {author} {\bibinfo {author} {\bibfnamefont {J.}~\bibnamefont
  {Bernon}}\ and\ \bibinfo {author} {\bibfnamefont {C.}~\bibnamefont {Smith}},\
  }\href {\doibase 10.1016/j.physletb.2016.03.068} {\bibfield  {journal}
  {\bibinfo  {journal} {Phys. Lett.}\ }\textbf {\bibinfo {volume} {B757}},\
  \bibinfo {pages} {148} (\bibinfo {year} {2016})},\ \Eprint
  {http://arxiv.org/abs/1512.06113} {arXiv:1512.06113} \BibitemShut {NoStop}%
\bibitem [{\citenamefont {An}\ \emph {et~al.}(2015)\citenamefont {An},
  \citenamefont {Cheung},\ and\ \citenamefont {Zhang}}]{An:2015cgp}%
  \BibitemOpen
  \bibfield  {author} {\bibinfo {author} {\bibfnamefont {H.}~\bibnamefont
  {An}}, \bibinfo {author} {\bibfnamefont {C.}~\bibnamefont {Cheung}}, \ and\
  \bibinfo {author} {\bibfnamefont {Y.}~\bibnamefont {Zhang}},\ }\href@noop {}
  {\  (\bibinfo {year} {2015})},\ \Eprint {http://arxiv.org/abs/1512.08378}
  {arXiv:1512.08378} \BibitemShut {NoStop}%
\bibitem [{\citenamefont {Knapen}\ \emph {et~al.}(2016)\citenamefont {Knapen},
  \citenamefont {Melia}, \citenamefont {Papucci},\ and\ \citenamefont
  {Zurek}}]{Knapen:2015dap}%
  \BibitemOpen
  \bibfield  {author} {\bibinfo {author} {\bibfnamefont {S.}~\bibnamefont
  {Knapen}}, \bibinfo {author} {\bibfnamefont {T.}~\bibnamefont {Melia}},
  \bibinfo {author} {\bibfnamefont {M.}~\bibnamefont {Papucci}}, \ and\
  \bibinfo {author} {\bibfnamefont {K.}~\bibnamefont {Zurek}},\ }\href
  {\doibase 10.1103/PhysRevD.93.075020} {\bibfield  {journal} {\bibinfo
  {journal} {Phys. Rev. D}\ }\textbf {\bibinfo {volume} {93}},\ \bibinfo
  {pages} {075020} (\bibinfo {year} {2016})},\ \Eprint
  {http://arxiv.org/abs/1512.04928} {arXiv:1512.04928} \BibitemShut {NoStop}%
\bibitem [{\citenamefont {Kim}\ \emph {et~al.}(2016)\citenamefont {Kim},
  \citenamefont {Reuter}, \citenamefont {Rolbiecki},\ and\ \citenamefont
  {Ruiz~de Austri}}]{Kim:2015ron}%
  \BibitemOpen
  \bibfield  {author} {\bibinfo {author} {\bibfnamefont {J.~S.}\ \bibnamefont
  {Kim}}, \bibinfo {author} {\bibfnamefont {J.}~\bibnamefont {Reuter}},
  \bibinfo {author} {\bibfnamefont {K.}~\bibnamefont {Rolbiecki}}, \ and\
  \bibinfo {author} {\bibfnamefont {R.}~\bibnamefont {Ruiz~de Austri}},\ }\href
  {\doibase 10.1016/j.physletb.2016.02.041} {\bibfield  {journal} {\bibinfo
  {journal} {Phys. Lett.}\ }\textbf {\bibinfo {volume} {B755}},\ \bibinfo
  {pages} {403} (\bibinfo {year} {2016})},\ \Eprint
  {http://arxiv.org/abs/1512.06083} {arXiv:1512.06083} \BibitemShut {NoStop}%
\bibitem [{\citenamefont {Cho}\ \emph {et~al.}(2016)\citenamefont {Cho},
  \citenamefont {Kim}, \citenamefont {Kong}, \citenamefont {Lim}, \citenamefont
  {Matchev}, \citenamefont {Park},\ and\ \citenamefont {Park}}]{Cho:2015nxy}%
  \BibitemOpen
  \bibfield  {author} {\bibinfo {author} {\bibfnamefont {W.~S.}\ \bibnamefont
  {Cho}}, \bibinfo {author} {\bibfnamefont {D.}~\bibnamefont {Kim}}, \bibinfo
  {author} {\bibfnamefont {K.}~\bibnamefont {Kong}}, \bibinfo {author}
  {\bibfnamefont {S.~H.}\ \bibnamefont {Lim}}, \bibinfo {author} {\bibfnamefont
  {K.~T.}\ \bibnamefont {Matchev}}, \bibinfo {author} {\bibfnamefont {J.-C.}\
  \bibnamefont {Park}}, \ and\ \bibinfo {author} {\bibfnamefont
  {M.}~\bibnamefont {Park}},\ }\href {\doibase 10.1103/PhysRevLett.116.151805}
  {\bibfield  {journal} {\bibinfo  {journal} {Phys. Rev. Lett.}\ }\textbf
  {\bibinfo {volume} {116}},\ \bibinfo {pages} {151805} (\bibinfo {year}
  {2016})},\ \Eprint {http://arxiv.org/abs/1512.06824} {arXiv:1512.06824
  [hep-ph]} \BibitemShut {NoStop}%
\bibitem [{\citenamefont {Altmannshofer}\ \emph {et~al.}(2016)\citenamefont
  {Altmannshofer}, \citenamefont {Galloway}, \citenamefont {Gori},
  \citenamefont {Kagan}, \citenamefont {Martin},\ and\ \citenamefont
  {Zupan}}]{Altmannshofer:2015xfo}%
  \BibitemOpen
  \bibfield  {author} {\bibinfo {author} {\bibfnamefont {W.}~\bibnamefont
  {Altmannshofer}}, \bibinfo {author} {\bibfnamefont {J.}~\bibnamefont
  {Galloway}}, \bibinfo {author} {\bibfnamefont {S.}~\bibnamefont {Gori}},
  \bibinfo {author} {\bibfnamefont {A.~L.}\ \bibnamefont {Kagan}}, \bibinfo
  {author} {\bibfnamefont {A.}~\bibnamefont {Martin}}, \ and\ \bibinfo {author}
  {\bibfnamefont {J.}~\bibnamefont {Zupan}},\ }\href {\doibase
  10.1103/PhysRevD.93.095015} {\bibfield  {journal} {\bibinfo  {journal} {Phys.
  Rev.}\ }\textbf {\bibinfo {volume} {D93}},\ \bibinfo {pages} {095015}
  (\bibinfo {year} {2016})},\ \Eprint {http://arxiv.org/abs/1512.07616}
  {arXiv:1512.07616 [hep-ph]} \BibitemShut {NoStop}%
\bibitem [{\citenamefont {Liu}\ \emph {et~al.}(2015)\citenamefont {Liu},
  \citenamefont {Wang},\ and\ \citenamefont {Xue}}]{Liu:2015yec}%
  \BibitemOpen
  \bibfield  {author} {\bibinfo {author} {\bibfnamefont {J.}~\bibnamefont
  {Liu}}, \bibinfo {author} {\bibfnamefont {X.-P.}\ \bibnamefont {Wang}}, \
  and\ \bibinfo {author} {\bibfnamefont {W.}~\bibnamefont {Xue}},\ }\href@noop
  {} {\  (\bibinfo {year} {2015})},\ \Eprint {http://arxiv.org/abs/1512.07885}
  {arXiv:1512.07885} \BibitemShut {NoStop}%
\bibitem [{\citenamefont {Franceschini}\ \emph
  {et~al.}(2016{\natexlab{b}})\citenamefont {Franceschini}, \citenamefont
  {Giudice}, \citenamefont {Kamenik}, \citenamefont {McCullough}, \citenamefont
  {Pomarol}, \citenamefont {Rattazzi}, \citenamefont {Redi}, \citenamefont
  {Riva}, \citenamefont {Strumia},\ and\ \citenamefont
  {Torre}}]{Franceschini:2015kwy}%
  \BibitemOpen
  \bibfield  {author} {\bibinfo {author} {\bibfnamefont {R.}~\bibnamefont
  {Franceschini}}, \bibinfo {author} {\bibfnamefont {G.~F.}\ \bibnamefont
  {Giudice}}, \bibinfo {author} {\bibfnamefont {J.~F.}\ \bibnamefont
  {Kamenik}}, \bibinfo {author} {\bibfnamefont {M.}~\bibnamefont {McCullough}},
  \bibinfo {author} {\bibfnamefont {A.}~\bibnamefont {Pomarol}}, \bibinfo
  {author} {\bibfnamefont {R.}~\bibnamefont {Rattazzi}}, \bibinfo {author}
  {\bibfnamefont {M.}~\bibnamefont {Redi}}, \bibinfo {author} {\bibfnamefont
  {F.}~\bibnamefont {Riva}}, \bibinfo {author} {\bibfnamefont {A.}~\bibnamefont
  {Strumia}}, \ and\ \bibinfo {author} {\bibfnamefont {R.}~\bibnamefont
  {Torre}},\ }\href {\doibase 10.1007/JHEP03(2016)144} {\bibfield  {journal}
  {\bibinfo  {journal} {JHEP}\ }\textbf {\bibinfo {volume} {03}},\ \bibinfo
  {pages} {144} (\bibinfo {year} {2016}{\natexlab{b}})},\ \Eprint
  {http://arxiv.org/abs/1512.04933} {arXiv:1512.04933} \BibitemShut {NoStop}%
\bibitem [{\citenamefont {Bauer}\ \emph {et~al.}(2000)\citenamefont {Bauer},
  \citenamefont {Fleming},\ and\ \citenamefont {Luke}}]{Bauer:2000ew}%
  \BibitemOpen
  \bibfield  {author} {\bibinfo {author} {\bibfnamefont {C.~W.}\ \bibnamefont
  {Bauer}}, \bibinfo {author} {\bibfnamefont {S.}~\bibnamefont {Fleming}}, \
  and\ \bibinfo {author} {\bibfnamefont {M.~E.}\ \bibnamefont {Luke}},\ }\href
  {\doibase 10.1103/PhysRevD.63.014006} {\bibfield  {journal} {\bibinfo
  {journal} {Phys. Rev. D}\ }\textbf {\bibinfo {volume} {63}},\ \bibinfo
  {pages} {014006} (\bibinfo {year} {2000})},\ \Eprint
  {http://arxiv.org/abs/hep-ph/0005275} {hep-ph/0005275} \BibitemShut {NoStop}%
\bibitem [{\citenamefont {Bauer}\ \emph {et~al.}(2001)\citenamefont {Bauer},
  \citenamefont {Fleming}, \citenamefont {Pirjol},\ and\ \citenamefont
  {Stewart}}]{Bauer:2000yr}%
  \BibitemOpen
  \bibfield  {author} {\bibinfo {author} {\bibfnamefont {C.~W.}\ \bibnamefont
  {Bauer}}, \bibinfo {author} {\bibfnamefont {S.}~\bibnamefont {Fleming}},
  \bibinfo {author} {\bibfnamefont {D.}~\bibnamefont {Pirjol}}, \ and\ \bibinfo
  {author} {\bibfnamefont {I.~W.}\ \bibnamefont {Stewart}},\ }\href {\doibase
  10.1103/PhysRevD.63.114020} {\bibfield  {journal} {\bibinfo  {journal} {Phys.
  Rev. D}\ }\textbf {\bibinfo {volume} {63}},\ \bibinfo {pages} {114020}
  (\bibinfo {year} {2001})},\ \Eprint {http://arxiv.org/abs/hep-ph/0011336}
  {hep-ph/0011336} \BibitemShut {NoStop}%
\bibitem [{\citenamefont {Bauer}\ and\ \citenamefont
  {Stewart}(2001)}]{Bauer:2001ct}%
  \BibitemOpen
  \bibfield  {author} {\bibinfo {author} {\bibfnamefont {C.~W.}\ \bibnamefont
  {Bauer}}\ and\ \bibinfo {author} {\bibfnamefont {I.~W.}\ \bibnamefont
  {Stewart}},\ }\href {\doibase 10.1016/S0370-2693(01)00902-9} {\bibfield
  {journal} {\bibinfo  {journal} {Phys. Lett.}\ }\textbf {\bibinfo {volume}
  {B516}},\ \bibinfo {pages} {134} (\bibinfo {year} {2001})},\ \Eprint
  {http://arxiv.org/abs/hep-ph/0107001} {hep-ph/0107001} \BibitemShut {NoStop}%
\bibitem [{\citenamefont {Bauer}\ \emph {et~al.}(2002)\citenamefont {Bauer},
  \citenamefont {Pirjol},\ and\ \citenamefont {Stewart}}]{Bauer:2001yt}%
  \BibitemOpen
  \bibfield  {author} {\bibinfo {author} {\bibfnamefont {C.~W.}\ \bibnamefont
  {Bauer}}, \bibinfo {author} {\bibfnamefont {D.}~\bibnamefont {Pirjol}}, \
  and\ \bibinfo {author} {\bibfnamefont {I.~W.}\ \bibnamefont {Stewart}},\
  }\href {\doibase 10.1103/PhysRevD.65.054022} {\bibfield  {journal} {\bibinfo
  {journal} {Phys. Rev. D}\ }\textbf {\bibinfo {volume} {65}},\ \bibinfo
  {pages} {054022} (\bibinfo {year} {2002})},\ \Eprint
  {http://arxiv.org/abs/hep-ph/0109045} {hep-ph/0109045} \BibitemShut {NoStop}%
\bibitem [{\citenamefont {Stewart}\ \emph {et~al.}(2010)\citenamefont
  {Stewart}, \citenamefont {Tackmann},\ and\ \citenamefont
  {Waalewijn}}]{Stewart:2009yx}%
  \BibitemOpen
  \bibfield  {author} {\bibinfo {author} {\bibfnamefont {I.~W.}\ \bibnamefont
  {Stewart}}, \bibinfo {author} {\bibfnamefont {F.~J.}\ \bibnamefont
  {Tackmann}}, \ and\ \bibinfo {author} {\bibfnamefont {W.~J.}\ \bibnamefont
  {Waalewijn}},\ }\href {\doibase 10.1103/PhysRevD.81.094035} {\bibfield
  {journal} {\bibinfo  {journal} {Phys. Rev. D}\ }\textbf {\bibinfo {volume}
  {81}},\ \bibinfo {pages} {094035} (\bibinfo {year} {2010})},\ \Eprint
  {http://arxiv.org/abs/0910.0467} {arXiv:0910.0467} \BibitemShut {NoStop}%
\bibitem [{\citenamefont {Berger}\ \emph {et~al.}(2011)\citenamefont {Berger},
  \citenamefont {Marcantonini}, \citenamefont {Stewart}, \citenamefont
  {Tackmann},\ and\ \citenamefont {Waalewijn}}]{Berger:2010xi}%
  \BibitemOpen
  \bibfield  {author} {\bibinfo {author} {\bibfnamefont {C.~F.}\ \bibnamefont
  {Berger}}, \bibinfo {author} {\bibfnamefont {C.}~\bibnamefont
  {Marcantonini}}, \bibinfo {author} {\bibfnamefont {I.~W.}\ \bibnamefont
  {Stewart}}, \bibinfo {author} {\bibfnamefont {F.~J.}\ \bibnamefont
  {Tackmann}}, \ and\ \bibinfo {author} {\bibfnamefont {W.~J.}\ \bibnamefont
  {Waalewijn}},\ }\href {\doibase 10.1007/JHEP04(2011)092} {\bibfield
  {journal} {\bibinfo  {journal} {JHEP}\ }\textbf {\bibinfo {volume} {04}},\
  \bibinfo {pages} {092} (\bibinfo {year} {2011})},\ \Eprint
  {http://arxiv.org/abs/1012.4480} {arXiv:1012.4480} \BibitemShut {NoStop}%
\bibitem [{\citenamefont {Moult}\ \emph {et~al.}(2016)\citenamefont {Moult},
  \citenamefont {Stewart}, \citenamefont {Tackmann},\ and\ \citenamefont
  {Waalewijn}}]{Moult:2015aoa}%
  \BibitemOpen
  \bibfield  {author} {\bibinfo {author} {\bibfnamefont {I.}~\bibnamefont
  {Moult}}, \bibinfo {author} {\bibfnamefont {I.~W.}\ \bibnamefont {Stewart}},
  \bibinfo {author} {\bibfnamefont {F.~J.}\ \bibnamefont {Tackmann}}, \ and\
  \bibinfo {author} {\bibfnamefont {W.~J.}\ \bibnamefont {Waalewijn}},\ }\href
  {\doibase 10.1103/PhysRevD.93.094003} {\bibfield  {journal} {\bibinfo
  {journal} {Phys. Rev. D}\ }\textbf {\bibinfo {volume} {93}},\ \bibinfo
  {pages} {094003} (\bibinfo {year} {2016})},\ \Eprint
  {http://arxiv.org/abs/1508.02397} {arXiv:1508.02397} \BibitemShut {NoStop}%
\bibitem [{\citenamefont {Moult}\ and\ \citenamefont
  {Stewart}(2014)}]{Moult:2014pja}%
  \BibitemOpen
  \bibfield  {author} {\bibinfo {author} {\bibfnamefont {I.}~\bibnamefont
  {Moult}}\ and\ \bibinfo {author} {\bibfnamefont {I.~W.}\ \bibnamefont
  {Stewart}},\ }\href {\doibase 10.1007/JHEP09(2014)129} {\bibfield  {journal}
  {\bibinfo  {journal} {JHEP}\ }\textbf {\bibinfo {volume} {09}},\ \bibinfo
  {pages} {129} (\bibinfo {year} {2014})},\ \Eprint
  {http://arxiv.org/abs/1405.5534} {arXiv:1405.5534} \BibitemShut {NoStop}%
\bibitem [{\citenamefont {Tackmann}\ \emph {et~al.}(2016)\citenamefont
  {Tackmann}, \citenamefont {Waalewijn},\ and\ \citenamefont
  {Zeune}}]{Tackmann:2016jyb}%
  \BibitemOpen
  \bibfield  {author} {\bibinfo {author} {\bibfnamefont {F.~J.}\ \bibnamefont
  {Tackmann}}, \bibinfo {author} {\bibfnamefont {W.~J.}\ \bibnamefont
  {Waalewijn}}, \ and\ \bibinfo {author} {\bibfnamefont {L.}~\bibnamefont
  {Zeune}},\ }\href {\doibase 10.1007/JHEP07(2016)119} {\bibfield  {journal}
  {\bibinfo  {journal} {JHEP}\ }\textbf {\bibinfo {volume} {07}},\ \bibinfo
  {pages} {119} (\bibinfo {year} {2016})},\ \Eprint
  {http://arxiv.org/abs/1603.03052} {arXiv:1603.03052 [hep-ph]} \BibitemShut
  {NoStop}%
\bibitem [{\citenamefont {Tackmann}\ \emph {et~al.}(2012)\citenamefont
  {Tackmann}, \citenamefont {Walsh},\ and\ \citenamefont
  {Zuberi}}]{Tackmann:2012bt}%
  \BibitemOpen
  \bibfield  {author} {\bibinfo {author} {\bibfnamefont {F.~J.}\ \bibnamefont
  {Tackmann}}, \bibinfo {author} {\bibfnamefont {J.~R.}\ \bibnamefont {Walsh}},
  \ and\ \bibinfo {author} {\bibfnamefont {S.}~\bibnamefont {Zuberi}},\ }\href
  {\doibase 10.1103/PhysRevD.86.053011} {\bibfield  {journal} {\bibinfo
  {journal} {Phys. Rev. D}\ }\textbf {\bibinfo {volume} {86}},\ \bibinfo
  {pages} {053011} (\bibinfo {year} {2012})},\ \Eprint
  {http://arxiv.org/abs/1206.4312} {arXiv:1206.4312} \BibitemShut {NoStop}%
\bibitem [{\citenamefont {Stewart}\ \emph {et~al.}(2014)\citenamefont
  {Stewart}, \citenamefont {Tackmann}, \citenamefont {Walsh},\ and\
  \citenamefont {Zuberi}}]{Stewart:2013faa}%
  \BibitemOpen
  \bibfield  {author} {\bibinfo {author} {\bibfnamefont {I.~W.}\ \bibnamefont
  {Stewart}}, \bibinfo {author} {\bibfnamefont {F.~J.}\ \bibnamefont
  {Tackmann}}, \bibinfo {author} {\bibfnamefont {J.~R.}\ \bibnamefont {Walsh}},
  \ and\ \bibinfo {author} {\bibfnamefont {S.}~\bibnamefont {Zuberi}},\ }\href
  {\doibase 10.1103/PhysRevD.89.054001} {\bibfield  {journal} {\bibinfo
  {journal} {Phys. Rev. D}\ }\textbf {\bibinfo {volume} {89}},\ \bibinfo
  {pages} {054001} (\bibinfo {year} {2014})},\ \Eprint
  {http://arxiv.org/abs/1307.1808} {arXiv:1307.1808} \BibitemShut {NoStop}%
\bibitem [{\citenamefont {Banfi}\ \emph {et~al.}(2012)\citenamefont {Banfi},
  \citenamefont {Monni}, \citenamefont {Salam},\ and\ \citenamefont
  {Zanderighi}}]{Banfi:2012jm}%
  \BibitemOpen
  \bibfield  {author} {\bibinfo {author} {\bibfnamefont {A.}~\bibnamefont
  {Banfi}}, \bibinfo {author} {\bibfnamefont {P.~F.}\ \bibnamefont {Monni}},
  \bibinfo {author} {\bibfnamefont {G.~P.}\ \bibnamefont {Salam}}, \ and\
  \bibinfo {author} {\bibfnamefont {G.}~\bibnamefont {Zanderighi}},\ }\href
  {\doibase 10.1103/PhysRevLett.109.202001} {\bibfield  {journal} {\bibinfo
  {journal} {Phys. Rev. Lett.}\ }\textbf {\bibinfo {volume} {109}},\ \bibinfo
  {pages} {202001} (\bibinfo {year} {2012})},\ \Eprint
  {http://arxiv.org/abs/1206.4998} {arXiv:1206.4998} \BibitemShut {NoStop}%
\bibitem [{\citenamefont {Becher}\ and\ \citenamefont
  {Neubert}(2012)}]{Becher:2012qa}%
  \BibitemOpen
  \bibfield  {author} {\bibinfo {author} {\bibfnamefont {T.}~\bibnamefont
  {Becher}}\ and\ \bibinfo {author} {\bibfnamefont {M.}~\bibnamefont
  {Neubert}},\ }\href {\doibase 10.1007/JHEP07(2012)108} {\bibfield  {journal}
  {\bibinfo  {journal} {JHEP}\ }\textbf {\bibinfo {volume} {07}},\ \bibinfo
  {pages} {108} (\bibinfo {year} {2012})},\ \Eprint
  {http://arxiv.org/abs/1205.3806} {arXiv:1205.3806} \BibitemShut {NoStop}%
\bibitem [{\citenamefont {Becher}\ \emph {et~al.}(2013)\citenamefont {Becher},
  \citenamefont {Neubert},\ and\ \citenamefont {Rothen}}]{Becher:2013xia}%
  \BibitemOpen
  \bibfield  {author} {\bibinfo {author} {\bibfnamefont {T.}~\bibnamefont
  {Becher}}, \bibinfo {author} {\bibfnamefont {M.}~\bibnamefont {Neubert}}, \
  and\ \bibinfo {author} {\bibfnamefont {L.}~\bibnamefont {Rothen}},\ }\href
  {\doibase 10.1007/JHEP10(2013)125} {\bibfield  {journal} {\bibinfo  {journal}
  {JHEP}\ }\textbf {\bibinfo {volume} {10}},\ \bibinfo {pages} {125} (\bibinfo
  {year} {2013})},\ \Eprint {http://arxiv.org/abs/1307.0025} {arXiv:1307.0025}
  \BibitemShut {NoStop}%
\bibitem [{\citenamefont {Banfi}\ \emph {et~al.}(2016)\citenamefont {Banfi},
  \citenamefont {Caola}, \citenamefont {Dreyer}, \citenamefont {Monni},
  \citenamefont {Salam}, \citenamefont {Zanderighi},\ and\ \citenamefont
  {Dulat}}]{Banfi:2015pju}%
  \BibitemOpen
  \bibfield  {author} {\bibinfo {author} {\bibfnamefont {A.}~\bibnamefont
  {Banfi}}, \bibinfo {author} {\bibfnamefont {F.}~\bibnamefont {Caola}},
  \bibinfo {author} {\bibfnamefont {F.~A.}\ \bibnamefont {Dreyer}}, \bibinfo
  {author} {\bibfnamefont {P.~F.}\ \bibnamefont {Monni}}, \bibinfo {author}
  {\bibfnamefont {G.~P.}\ \bibnamefont {Salam}}, \bibinfo {author}
  {\bibfnamefont {G.}~\bibnamefont {Zanderighi}}, \ and\ \bibinfo {author}
  {\bibfnamefont {F.}~\bibnamefont {Dulat}},\ }\href {\doibase
  10.1007/JHEP04(2016)049} {\bibfield  {journal} {\bibinfo  {journal} {JHEP}\
  }\textbf {\bibinfo {volume} {04}},\ \bibinfo {pages} {049} (\bibinfo {year}
  {2016})},\ \Eprint {http://arxiv.org/abs/1511.02886} {arXiv:1511.02886}
  \BibitemShut {NoStop}%
\bibitem [{\citenamefont {Harlander}\ \emph {et~al.}(2013)\citenamefont
  {Harlander}, \citenamefont {Liebler},\ and\ \citenamefont
  {Mantler}}]{Harlander:2012pb}%
  \BibitemOpen
  \bibfield  {author} {\bibinfo {author} {\bibfnamefont {R.~V.}\ \bibnamefont
  {Harlander}}, \bibinfo {author} {\bibfnamefont {S.}~\bibnamefont {Liebler}},
  \ and\ \bibinfo {author} {\bibfnamefont {H.}~\bibnamefont {Mantler}},\ }\href
  {\doibase 10.1016/j.cpc.2013.02.006} {\bibfield  {journal} {\bibinfo
  {journal} {Comput. Phys. Commun.}\ }\textbf {\bibinfo {volume} {184}},\
  \bibinfo {pages} {1605} (\bibinfo {year} {2013})},\ \Eprint
  {http://arxiv.org/abs/1212.3249} {arXiv:1212.3249} \BibitemShut {NoStop}%
\bibitem [{\citenamefont {Harlander}\ \emph {et~al.}(2016)\citenamefont
  {Harlander}, \citenamefont {Liebler},\ and\ \citenamefont
  {Mantler}}]{Harlander:2016hcx}%
  \BibitemOpen
  \bibfield  {author} {\bibinfo {author} {\bibfnamefont {R.~V.}\ \bibnamefont
  {Harlander}}, \bibinfo {author} {\bibfnamefont {S.}~\bibnamefont {Liebler}},
  \ and\ \bibinfo {author} {\bibfnamefont {H.}~\bibnamefont {Mantler}},\
  }\href@noop {} {\  (\bibinfo {year} {2016})},\ \Eprint
  {http://arxiv.org/abs/1605.03190} {arXiv:1605.03190} \BibitemShut {NoStop}%
\bibitem [{\citenamefont {Harlander}(2016)}]{Harlander:2015xur}%
  \BibitemOpen
  \bibfield  {author} {\bibinfo {author} {\bibfnamefont {R.~V.}\ \bibnamefont
  {Harlander}},\ }\href {\doibase 10.1140/epjc/s10052-016-4093-x} {\bibfield
  {journal} {\bibinfo  {journal} {Eur. Phys. J.}\ }\textbf {\bibinfo {volume}
  {C76}},\ \bibinfo {pages} {252} (\bibinfo {year} {2016})},\ \Eprint
  {http://arxiv.org/abs/1512.04901} {arXiv:1512.04901} \BibitemShut {NoStop}%
\bibitem [{\citenamefont {Harlander}\ and\ \citenamefont
  {Kilgore}(2002)}]{Harlander:2002wh}%
  \BibitemOpen
  \bibfield  {author} {\bibinfo {author} {\bibfnamefont {R.~V.}\ \bibnamefont
  {Harlander}}\ and\ \bibinfo {author} {\bibfnamefont {W.~B.}\ \bibnamefont
  {Kilgore}},\ }\href {\doibase 10.1103/PhysRevLett.88.201801} {\bibfield
  {journal} {\bibinfo  {journal} {Phys. Rev. Lett.}\ }\textbf {\bibinfo
  {volume} {88}},\ \bibinfo {pages} {201801} (\bibinfo {year} {2002})},\
  \Eprint {http://arxiv.org/abs/hep-ph/0201206} {hep-ph/0201206} \BibitemShut
  {NoStop}%
\bibitem [{\citenamefont {Harlander}\ \emph {et~al.}(2010)\citenamefont
  {Harlander}, \citenamefont {Ozeren},\ and\ \citenamefont
  {Wiesemann}}]{Harlander:2010cz}%
  \BibitemOpen
  \bibfield  {author} {\bibinfo {author} {\bibfnamefont {R.~V.}\ \bibnamefont
  {Harlander}}, \bibinfo {author} {\bibfnamefont {K.~J.}\ \bibnamefont
  {Ozeren}}, \ and\ \bibinfo {author} {\bibfnamefont {M.}~\bibnamefont
  {Wiesemann}},\ }\href {\doibase 10.1016/j.physletb.2010.08.038} {\bibfield
  {journal} {\bibinfo  {journal} {Phys. Lett.}\ }\textbf {\bibinfo {volume}
  {B693}},\ \bibinfo {pages} {269} (\bibinfo {year} {2010})},\ \Eprint
  {http://arxiv.org/abs/1007.5411} {arXiv:1007.5411} \BibitemShut {NoStop}%
\bibitem [{\citenamefont {Campbell}\ and\ \citenamefont
  {Ellis}(1999)}]{Campbell:1999ah}%
  \BibitemOpen
  \bibfield  {author} {\bibinfo {author} {\bibfnamefont {J.~M.}\ \bibnamefont
  {Campbell}}\ and\ \bibinfo {author} {\bibfnamefont {R.~K.}\ \bibnamefont
  {Ellis}},\ }\href {\doibase 10.1103/PhysRevD.60.113006} {\bibfield  {journal}
  {\bibinfo  {journal} {Phys. Rev. D}\ }\textbf {\bibinfo {volume} {60}},\
  \bibinfo {pages} {113006} (\bibinfo {year} {1999})},\ \Eprint
  {http://arxiv.org/abs/hep-ph/9905386} {arXiv:hep-ph/9905386} \BibitemShut
  {NoStop}%
\bibitem [{\citenamefont {Campbell}\ \emph {et~al.}(2011)\citenamefont
  {Campbell}, \citenamefont {Ellis},\ and\ \citenamefont
  {Williams}}]{Campbell:2011bn}%
  \BibitemOpen
  \bibfield  {author} {\bibinfo {author} {\bibfnamefont {J.~M.}\ \bibnamefont
  {Campbell}}, \bibinfo {author} {\bibfnamefont {R.~K.}\ \bibnamefont {Ellis}},
  \ and\ \bibinfo {author} {\bibfnamefont {C.}~\bibnamefont {Williams}},\
  }\href {\doibase 10.1007/JHEP07(2011)018} {\bibfield  {journal} {\bibinfo
  {journal} {JHEP}\ }\textbf {\bibinfo {volume} {07}},\ \bibinfo {pages} {018}
  (\bibinfo {year} {2011})},\ \Eprint {http://arxiv.org/abs/1105.0020}
  {arXiv:1105.0020} \BibitemShut {NoStop}%
\bibitem [{\citenamefont {Stewart}\ and\ \citenamefont
  {Tackmann}(2012)}]{Stewart:2011cf}%
  \BibitemOpen
  \bibfield  {author} {\bibinfo {author} {\bibfnamefont {I.~W.}\ \bibnamefont
  {Stewart}}\ and\ \bibinfo {author} {\bibfnamefont {F.~J.}\ \bibnamefont
  {Tackmann}},\ }\href {\doibase 10.1103/PhysRevD.85.034011} {\bibfield
  {journal} {\bibinfo  {journal} {Phys. Rev. D}\ }\textbf {\bibinfo {volume}
  {85}},\ \bibinfo {pages} {034011} (\bibinfo {year} {2012})},\ \Eprint
  {http://arxiv.org/abs/1107.2117} {arXiv:1107.2117} \BibitemShut {NoStop}%
\bibitem [{\citenamefont {Ligeti}\ \emph {et~al.}(2008)\citenamefont {Ligeti},
  \citenamefont {Stewart},\ and\ \citenamefont {Tackmann}}]{Ligeti:2008ac}%
  \BibitemOpen
  \bibfield  {author} {\bibinfo {author} {\bibfnamefont {Z.}~\bibnamefont
  {Ligeti}}, \bibinfo {author} {\bibfnamefont {I.~W.}\ \bibnamefont {Stewart}},
  \ and\ \bibinfo {author} {\bibfnamefont {F.~J.}\ \bibnamefont {Tackmann}},\
  }\href {\doibase 10.1103/PhysRevD.78.114014} {\bibfield  {journal} {\bibinfo
  {journal} {Phys. Rev. D}\ }\textbf {\bibinfo {volume} {78}},\ \bibinfo
  {pages} {114014} (\bibinfo {year} {2008})},\ \Eprint
  {http://arxiv.org/abs/0807.1926} {arXiv:0807.1926} \BibitemShut {NoStop}%
\bibitem [{\citenamefont {Abbate}\ \emph {et~al.}(2011)\citenamefont {Abbate},
  \citenamefont {Fickinger}, \citenamefont {Hoang}, \citenamefont {Mateu},\
  and\ \citenamefont {Stewart}}]{Abbate:2010xh}%
  \BibitemOpen
  \bibfield  {author} {\bibinfo {author} {\bibfnamefont {R.}~\bibnamefont
  {Abbate}}, \bibinfo {author} {\bibfnamefont {M.}~\bibnamefont {Fickinger}},
  \bibinfo {author} {\bibfnamefont {A.~H.}\ \bibnamefont {Hoang}}, \bibinfo
  {author} {\bibfnamefont {V.}~\bibnamefont {Mateu}}, \ and\ \bibinfo {author}
  {\bibfnamefont {I.~W.}\ \bibnamefont {Stewart}},\ }\href {\doibase
  10.1103/PhysRevD.83.074021} {\bibfield  {journal} {\bibinfo  {journal} {Phys.
  Rev. D}\ }\textbf {\bibinfo {volume} {83}},\ \bibinfo {pages} {074021}
  (\bibinfo {year} {2011})},\ \Eprint {http://arxiv.org/abs/1006.3080}
  {arXiv:1006.3080} \BibitemShut {NoStop}%
\bibitem [{\citenamefont {Harland-Lang}\ \emph {et~al.}(2015)\citenamefont
  {Harland-Lang}, \citenamefont {Martin}, \citenamefont {Motylinski},\ and\
  \citenamefont {Thorne}}]{Harland-Lang:2014zoa}%
  \BibitemOpen
  \bibfield  {author} {\bibinfo {author} {\bibfnamefont {L.~A.}\ \bibnamefont
  {Harland-Lang}}, \bibinfo {author} {\bibfnamefont {A.~D.}\ \bibnamefont
  {Martin}}, \bibinfo {author} {\bibfnamefont {P.}~\bibnamefont {Motylinski}},
  \ and\ \bibinfo {author} {\bibfnamefont {R.~S.}\ \bibnamefont {Thorne}},\
  }\href {\doibase 10.1140/epjc/s10052-015-3397-6} {\bibfield  {journal}
  {\bibinfo  {journal} {Eur. Phys. J.}\ }\textbf {\bibinfo {volume} {C75}},\
  \bibinfo {pages} {204} (\bibinfo {year} {2015})},\ \Eprint
  {http://arxiv.org/abs/1412.3989} {arXiv:1412.3989} \BibitemShut {NoStop}%
\bibitem [{\citenamefont {Cs{\'a}ki}\ \emph {et~al.}(2016)\citenamefont
  {Cs{\'a}ki}, \citenamefont {Hubisz},\ and\ \citenamefont
  {Terning}}]{Csaki:2015vek}%
  \BibitemOpen
  \bibfield  {author} {\bibinfo {author} {\bibfnamefont {C.}~\bibnamefont
  {Cs{\'a}ki}}, \bibinfo {author} {\bibfnamefont {J.}~\bibnamefont {Hubisz}}, \
  and\ \bibinfo {author} {\bibfnamefont {J.}~\bibnamefont {Terning}},\ }\href
  {\doibase 10.1103/PhysRevD.93.035002} {\bibfield  {journal} {\bibinfo
  {journal} {Phys. Rev. D}\ }\textbf {\bibinfo {volume} {93}},\ \bibinfo
  {pages} {035002} (\bibinfo {year} {2016})},\ \Eprint
  {http://arxiv.org/abs/1512.05776} {arXiv:1512.05776} \BibitemShut {NoStop}%
\bibitem [{\citenamefont {Fichet}\ \emph {et~al.}(2016)\citenamefont {Fichet},
  \citenamefont {von Gersdorff},\ and\ \citenamefont {Royon}}]{Fichet:2015vvy}%
  \BibitemOpen
  \bibfield  {author} {\bibinfo {author} {\bibfnamefont {S.}~\bibnamefont
  {Fichet}}, \bibinfo {author} {\bibfnamefont {G.}~\bibnamefont {von
  Gersdorff}}, \ and\ \bibinfo {author} {\bibfnamefont {C.}~\bibnamefont
  {Royon}},\ }\href {\doibase 10.1103/PhysRevD.93.075031} {\bibfield  {journal}
  {\bibinfo  {journal} {Phys. Rev. D}\ }\textbf {\bibinfo {volume} {93}},\
  \bibinfo {pages} {075031} (\bibinfo {year} {2016})},\ \Eprint
  {http://arxiv.org/abs/1512.05751} {arXiv:1512.05751} \BibitemShut {NoStop}%
\bibitem [{\citenamefont {Harland-Lang}\ \emph {et~al.}(2016)\citenamefont
  {Harland-Lang}, \citenamefont {Khoze},\ and\ \citenamefont
  {Ryskin}}]{Harland-Lang:2016qjy}%
  \BibitemOpen
  \bibfield  {author} {\bibinfo {author} {\bibfnamefont {L.~A.}\ \bibnamefont
  {Harland-Lang}}, \bibinfo {author} {\bibfnamefont {V.~A.}\ \bibnamefont
  {Khoze}}, \ and\ \bibinfo {author} {\bibfnamefont {M.~G.}\ \bibnamefont
  {Ryskin}},\ }\href {\doibase 10.1007/JHEP03(2016)182} {\bibfield  {journal}
  {\bibinfo  {journal} {JHEP}\ }\textbf {\bibinfo {volume} {03}},\ \bibinfo
  {pages} {182} (\bibinfo {year} {2016})},\ \Eprint
  {http://arxiv.org/abs/1601.07187} {arXiv:1601.07187} \BibitemShut {NoStop}%
\bibitem [{\citenamefont {Abel}\ and\ \citenamefont
  {Khoze}(2016)}]{Abel:2016pyc}%
  \BibitemOpen
  \bibfield  {author} {\bibinfo {author} {\bibfnamefont {S.}~\bibnamefont
  {Abel}}\ and\ \bibinfo {author} {\bibfnamefont {V.~V.}\ \bibnamefont
  {Khoze}},\ }\href {\doibase 10.1007/JHEP05(2016)063} {\bibfield  {journal}
  {\bibinfo  {journal} {JHEP}\ }\textbf {\bibinfo {volume} {05}},\ \bibinfo
  {pages} {063} (\bibinfo {year} {2016})},\ \Eprint
  {http://arxiv.org/abs/1601.07167} {arXiv:1601.07167} \BibitemShut {NoStop}%
\bibitem [{\citenamefont {Martin}\ and\ \citenamefont
  {Ryskin}(2016)}]{Martin:2016byf}%
  \BibitemOpen
  \bibfield  {author} {\bibinfo {author} {\bibfnamefont {A.~D.}\ \bibnamefont
  {Martin}}\ and\ \bibinfo {author} {\bibfnamefont {M.~G.}\ \bibnamefont
  {Ryskin}},\ }\href {\doibase 10.1088/0954-3899/43/4/04LT02} {\bibfield
  {journal} {\bibinfo  {journal} {J. Phys.}\ }\textbf {\bibinfo {volume}
  {G43}},\ \bibinfo {pages} {04LT02} (\bibinfo {year} {2016})},\ \Eprint
  {http://arxiv.org/abs/1601.07774} {arXiv:1601.07774} \BibitemShut {NoStop}%
\bibitem [{\citenamefont {Cynolter}\ \emph {et~al.}(2016)\citenamefont
  {Cynolter}, \citenamefont {Kovács},\ and\ \citenamefont
  {Lendvai}}]{Cynolter:2016jxv}%
  \BibitemOpen
  \bibfield  {author} {\bibinfo {author} {\bibfnamefont {G.}~\bibnamefont
  {Cynolter}}, \bibinfo {author} {\bibfnamefont {J.~.}\ \bibnamefont
  {Kovács}}, \ and\ \bibinfo {author} {\bibfnamefont {E.}~\bibnamefont
  {Lendvai}},\ }\href {\doibase 10.1142/S0217732316501339} {\bibfield
  {journal} {\bibinfo  {journal} {Mod. Phys. Lett.}\ }\textbf {\bibinfo
  {volume} {A31}},\ \bibinfo {pages} {1650133} (\bibinfo {year} {2016})},\
  \Eprint {http://arxiv.org/abs/1604.01008} {arXiv:1604.01008 [hep-ph]}
  \BibitemShut {NoStop}%
\bibitem [{\citenamefont {Csáki}\ \emph {et~al.}(2016)\citenamefont {Csáki},
  \citenamefont {Hubisz}, \citenamefont {Lombardo},\ and\ \citenamefont
  {Terning}}]{Csaki:2016raa}%
  \BibitemOpen
  \bibfield  {author} {\bibinfo {author} {\bibfnamefont {C.}~\bibnamefont
  {Csáki}}, \bibinfo {author} {\bibfnamefont {J.}~\bibnamefont {Hubisz}},
  \bibinfo {author} {\bibfnamefont {S.}~\bibnamefont {Lombardo}}, \ and\
  \bibinfo {author} {\bibfnamefont {J.}~\bibnamefont {Terning}},\ }\href
  {\doibase 10.1103/PhysRevD.93.095020} {\bibfield  {journal} {\bibinfo
  {journal} {Phys. Rev.}\ }\textbf {\bibinfo {volume} {D93}},\ \bibinfo {pages}
  {095020} (\bibinfo {year} {2016})},\ \Eprint
  {http://arxiv.org/abs/1601.00638} {arXiv:1601.00638 [hep-ph]} \BibitemShut
  {NoStop}%
\bibitem [{\citenamefont {Molinaro}\ \emph {et~al.}(2016)\citenamefont
  {Molinaro}, \citenamefont {Sannino},\ and\ \citenamefont
  {Vignaroli}}]{Molinaro:2016oix}%
  \BibitemOpen
  \bibfield  {author} {\bibinfo {author} {\bibfnamefont {E.}~\bibnamefont
  {Molinaro}}, \bibinfo {author} {\bibfnamefont {F.}~\bibnamefont {Sannino}}, \
  and\ \bibinfo {author} {\bibfnamefont {N.}~\bibnamefont {Vignaroli}},\ }\href
  {\doibase 10.1016/j.nuclphysb.2016.07.032} {\bibfield  {journal} {\bibinfo
  {journal} {Nucl. Phys.}\ }\textbf {\bibinfo {volume} {B911}},\ \bibinfo
  {pages} {106} (\bibinfo {year} {2016})},\ \Eprint
  {http://arxiv.org/abs/1602.07574} {arXiv:1602.07574 [hep-ph]} \BibitemShut
  {NoStop}%
\end{thebibliography}%

\end{document}